\documentclass[preprint,authoryear,11pt,3p]{elsarticle}

\usepackage{amssymb}
\usepackage{tikz}
\usepackage{graphicx}
\usepackage{caption}
\usepackage{mathtools}
\usepackage{amsmath}
\usepackage{multirow}
\usepackage{colortbl}
\usepackage{color}
\usepackage{tabulary}
\usepackage{bm}
\usepackage{eqparbox}
\usepackage{tabularx}
\usepackage[utf8]{inputenc}
\usepackage{booktabs}
\usepackage{kpfonts}
\usepackage{rotating}
\usepackage{enumitem}
\usepackage{makecell}
\usepackage{float}
\makeatletter
\AddToHook{env/tabular/begin}{\let\input\@@input}
\makeatother
\usepackage{amsthm}

\begin{document}
\begin{frontmatter}

\title{Scaleable Dynamic Forecast Reconciliation}

\author[label3,label1]{Ross Hollyman}
\ead{rah98@bath.ac.uk}
\author[label1]{Fotios Petropoulos\corref{cor1}}
\cortext[cor1]{Corresponding author: f.petropoulos@bath.ac.uk; fotios@bath.edu}
\ead{f.petropoulos@bath.ac.uk}
\author[label2]{Michael E. Tipping}
\ead{mt821@bath.ac.uk}
\address[label3]{School of Management, University of Bath, United Kingdom}
\address[label1]{School of Management, University of Bath, United Kingdom}
\address[label2]{Department of Computer Science, University of Bath, United Kingdom}
\date{March 2024}

\begin{abstract}
We introduce a dynamic approach to probabilistic forecast reconciliation at scale. Our model differs from the existing literature in this area in several important ways. Firstly we explicitly allow the weights allocated to the base forecasts in forming the combined, reconciled forecasts to vary over time. Secondly we drop the assumption, near ubiquitous in the literature,  that in-sample base forecasts are appropriate for determining these weights, and use out of sample forecasts instead. Most existing probabilistic reconciliation approaches rely on time consuming sampling based techniques, and therefore do not scale well (or at all) to large data sets. We address this problem in two main ways, firstly by utilising a closed from estimator of covariance structure appropriate to hierarchical forecasting problems, and secondly by decomposing large hierarchies in to components which can be reconciled separately.
\end{abstract}
\end{frontmatter}

\section{Introduction}

Forecast reconciliation is a multivariate forecasting technique, usually deployed on large (and potentially very large) collections of time series. The basic idea is that in the rather common situation where interest lies in several aggregates of a set of time series, in addition to the constituent themselves, forecast accuracy can be improved by combining the information in independently generated forecasts of the components and the aggregates. 

For example, in the simple case of a system of three time series, $X$, $Y$ and $T$, such that $T = X + Y$, the idea is that instead of forecasting $X$ and $Y$ separately and summing the resultant forecasts to forecast $T$, we forecast $X$, $Y$ \emph{and} $T$ independently and then \emph{reconcile} or adjust the resultant forecasts so that they conform to the relationship $T = X + Y$. In theory and in practice, the reconciliation based approach tends to be more somewhat more accurate. Aggregation or averaging of data has of course been used to increase the clarity of a signal for time immemorial, and indeed aggregated time series tend to be 'smoother' and easier to forecast, with, for instance, clearer patterns of trend and seasonality then noisier component series. Forecast reconciliation techniques allow the information in the forecasts of these aggregate series to shared across the constituent series.

The idea is clear in the context of time series which can be represented in a hierarchy, such as product sales for a commercial organisation, arranged perhaps by geographical region and product category. Interest may lie in forecasts at a product/geography level, at total product level, for the company as a whole, or some combination of all three. Indeed, in large corporate settings, forecasts at each level of the hierarchy may be produced by different teams in different departments \cite{Seaman2018-zv}. Forecast reconciliation techniques synthesise these independently produced forecasts so that they 'add up', and the power of forecast combination (\cite{Timmermann2006-ok}, \cite{Wang2023-ba}, \cite{Hollyman2021-ek}) generally leads to more accurate forecasts. Time series arranged in this way are somewhat naturally grouped in to 'aggregates' ($Z$ in our example above) and or 'bottom level' (in a hierarchical sense) or dis-aggregate series ($X$ and $Y$). We would argue that the obvious name for such series is 'base' series (especially as they are often written as a vector $\mathbf{b}_t$). Unfortunately, the term 'base' has historically been used in the literature to refer to the sets of forecasts which form the inputs to the reconciliation process. This sits somewhat uncomfortably with our approach, so we term these as 'exogenous' forecasts (as in exogenous to the reconciliation algorithms we describe) and use the word 'base' to refer to bottom level or dis-aggregate time series. As pointed out in \cite{Panagiotelis2021-sm} and \cite{Di_Fonzo2022-be}, reconciliation methods can be used where there is no natural hierarchical representation of the series, so the term base for these constituent series seems more appropriate.

The same techniques can deployed in other contexts. For example, when forecasting corporate profitability, a common approach would be to produce predictions of sales and costs and then derive forecasts of profit as a difference between the two. The  forecast reconciliation approach would instead produce separate forecasts of sales, costs and profits, and apply a reconciliation process to produce a synthesis of information from each separate model. In a recent paper (\cite{Di_Fonzo2022-be}), demonstrate a more general 'zero-constrained' formation of the problem applicable to a very wide variety of linearly constrained forecasting problems such as these.

One of the advantages of the hierarchical framework is that it is agnostic to the methodology used to forecast individual series in the collection; a wide variety of systematic and judgemental forecasting approaches can be utilised to do so.
 
Recent years have seen many applications of hierarchical forecasting approaches to new data sets, and have suggested a number of enhancements to the popular and well cited trace minimisation (MINT) models of \cite{Wickramasuriya2019-av}. Additionally, several authors, starting with \cite{Athanasopoulos2017-nw} extended the basic approach to incorporate information from forecasts of data aggregated over time (for example; forecasts of quarterly aggregates based on data originally observed at monthly frequency). These temporal approaches have been combined with cross sectional approaches to generate 'cross-temporal' models (\cite{Kourentzes2019-fy}).

The literature has also seen a renewed focus on probabilistic approaches to the problem, with most authors (for example \cite{Panagiotelis2023-tg} and \cite{GIROLIMETTO2023}) adapting approaches based on optimal approach of \cite{Wickramasuriya2019-av}. We refer the interested reader to the recent review of the literature in \cite{Athanasopoulos2024-zf} for further details of recent developments in the field.

Although attracting significant academic attention, several barriers remain to implementing reconciliation based approaches in practice, and especially for large systems where benefits are potentially particularly valuable; models which perform well on small to moderately sized datasets common in academic work fail to scale at all to these real world problems.

The reconciliation process is inherently computationally intensive. Fundamental to most approaches is some estimate of the variance / covariance matrix of the time series which make up the system. Clearly, the size of this matrix grows with the square of the number of observations. As systems increase in size, the computational load required to perform the reconciliation explodes. Consider what might be considered a relatively large system by academic standards - comprising 1,000 time series - using 32 bit numerical precision, Python requires 0.4GB of memory to store the variance covariance matrix. If the number of series increases to 100,000, then the corresponding memory requirement becomes 40GB, before any calculations are performed using the matrix. These numbers increase if higher numerical precision is deemed necessary. As large retailers often have many thousands of product lines, and many hundreds of stores \cite{Seaman2018-zv}, resulting in billions of store/product combinations, the situation rapidly becomes computationally infeasible.

Aside from computational load, covariance estimation at scale - where the number of time series often greatly exceeds the number of observations, is statistically challenging, even when data is plentiful. A further and related challenge relates to 'the curse of dimensionality' - a ubiquitous issue in multivariate analysis; the situation where many more cross-sectional units of data are available than there is time-series history for estimation. Accepting that reality that covariance structures can and do change over time reinforces the pressing need to confront this issue head on, and not just wish it away, for example simply by presenting models which assume that base time series are uncorrelated. We discuss these challenges in much more detail below and outline our suggested solution. 

The practical problems we describe are exacerbated if we seek probabilistic rather than point-forecast solutions. \cite{Wickramasuriya2019-av} present closed form estimators for forecast errors, but these depend on the same estimates of the base forecast error variance / covariance matrix described above, and disregard uncertainty in the underlying forecasts. Models which attempt to account for this key source of uncertainty, whether fully
Bayesian (\cite{hollyman2022hierarchies}) or otherwise (\cite{Panagiotelis2023-tg}, \cite{GIROLIMETTO2023}), tend to rely on sampling based approaches. Although  modern computational tools can be very efficient, this severely limits scalability to larger problems involving thousands of time series. 

We turn now to the simplifying assumptions common to most recent work in hierarchical forecasting, relating in particular to the calibration of the base forecasts, which may be problematic for several reasons.

Firstly the common approach is to take the exogenous forecasts \emph{taken as a whole} at face value in the sense that reconciliation process allocates weights (summing to 1 across base forecasts) so allocation decisions are made purely amongst base forecasts, in a manner similar to \cite{Bates1969-wz} forecast combination. Such allocations are made purely on the basis of \emph{in-sample} forecast accuracy (assuming squared error loss). The assumption here is that out-of-sample predictive ability of each forecast is the same as (or at least proportional to) one step ahead in-sample forecast accuracy, and that the out-of-sample error covariance structure of the forecasts sufficiently mirrors that of the in-sample forecasts. Where a variety of forecasting approaches are used in the generation of the exogenous forecasts (which is likely to be beneficial) this is likely to bias towards selection of over-parameterised models which we know tend to perform relatively poorly out of sample.

Secondly this assumption is usually extended to multi-step ahead forecasts. The estimated, one step ahead, in-sample forecast error covariance matrix is used to reconcile multi-step ahead forecasts, again the assumption being that the multi step ahead forecast error variance / covariance structure is appropriate at all forecast horizons.

Finally, the standard approach usually relies on fixed estimates of forecast error covariance, usually estimated over rolling rolling windows. In practice in business and economic circumstances, covariance between series and between forecasts and outcomes are likely to vary over time, and such processes can be slow to adapt to changes in circumstances.

Recent literature on hierarchical forecasting has tended to neglect an earlier Bayesian approach to forecasting with aggregates, outlined in \cite{West1999-am}. This approach differs fundamentally from MinT type approaches in that it assumes the availability of a prior estimate of the mean and variance / covariance matrix of the base series, and proceeds to update these quantities in the light of independent or exogenous forecast information. In this paper we adopt the basic approach of these authors, as doing so gives rise to considerable flexibility in estimation which allows many of the scalability issues of MINT style approaches to be overcome.

The focus of this paper is an empirical, rather than theoretical one, and we focus on a set of simplifications and approximations which enable us to reconcile very large sets of forecasts. We defer to the work of others, in particular \cite{Di_Fonzo2022-be} for rigorous statistical analysis of the theory underlying the reconciliation process. We proceeds as follows. In the next section we describe at a high level the fundamental components of our approach, and how these combine to produce reconciled forecasts at scale. The following section describes the algorithm in more technical detail. In section 4 we illustrate the algorithm via an application to the Australian Tourism data (comprising 525 separate time series) with which many practitioners in the field will be familiar. We then provide a second empirical example, applying our model to the M5 competition dataset (\cite{Makridakis2022-ur}) which comprises approximately 42,000 time series of daily data measured over approximately 5 years. The results of this exercise are described in section 5. Section 6 concludes.

\subsection{Notation}

We adhere closely as possible to the revised notation for HF adopted in \cite{GIROLIMETTO2023}. That is we let $\mathbf{b}_t$ represent an $n_b$ vector of what were initially referred to as the 'bottom level' (as above we prefer the term 'base') series in a hierarchy at time t. 
 We denote the $n_a$ aggregate series as $\mathbf{a}_t$ so that the entire collection of series is defined as the $n$ vector:

\begin{align*}
    \mathbf{y}_{t}=\left[\begin{matrix}\mathbf{a}_{t} \\ \mathbf{b}_{t}\end{matrix}\right]
\end{align*}

The same authors then use the letter $\mathbf{C}$ to denote a matrix embedding a collection of linear combinations of the base series $\mathbf{b}_t$ so that 

\begin{align*}
    \mathbf{a}_t = \mathbf{C} \mathbf{b}_t 
\end{align*}

and the HF summing matrix $\mathbf{S}$ is

\begin{align*}
    \mathbf{S} =\left[\begin{matrix}\mathbf{C} \\ \mathbf{I}_{n_b}\end{matrix}\right]
\end{align*}

We note that the matrix $\mathbf{C}$ here is different to the matrix $\mathbf{C}$ traditionally used to denote the posterior state variance in the standard notation for time series Dynamic Linear Models (DLMs \cite{West1999-am}) which we use extensively in this paper. 

We refer to the rows of matrix $\mathbf{C}$ as individual vectors $\{\mathbf{c}_i:i=1...n_a\}$.  In this paper we will essentially be concerned with combining a number of different forecasts of the vector $\mathbf{y}_t$. In the hierarchical forecasting literature it is common to refer to the exogenous forecasts of  $\mathbf{y}_t$ as $\hat{\mathbf{y}}_t$, and the reconciled forecasts of $\mathbf{y}_t$ as $\tilde{\mathbf{y}}_t = \mathbf{S}\tilde{\mathbf{b}}_t$. We shall additionally refer to the set of 'prior' forecasts of $\mathbf{y}_t$ as $\bar{\mathbf{y}}_t = \mathbf{S}\bar{\mathbf{b}}_t$, in the sense that we assume that the prior is established form the long run behaviour of the series, before exogenous forecasts are observed. 

In many hierarchical settings, the time series represent integer counts, and the base series especially are often highly intermittent, with many zero observations. Obviously, the standard analysis describes how distributions become approximately Gaussian under aggregation. For the purposes of this paper, we abstract away from this, and proceed in the spirit of 'linear bayes' analysis of \cite{Goldstein2007-it} and focus exclusively on the first two moments of the forecast distributions so that the notation $x \sim [a,B]$ should be read as '$x$ is distributed with mean $a$ and Variance $B$'. We leave the adoption of our techniques to intermittent time series to a future paper.

\section{Dynamic Forecast Reconciliation}

A stylised description of the most common approach to hierarchical forecasting proceeds broadly as follows:

\subsection{Stylised Approach}

\begin{enumerate}
    \item Estimate one exogenous forecast for each of the time series (aggregates and base series) in the collection
    \item Estimate the (in-sample) error variance / covariance matrix based on residuals from the exogenous forecasts generated in (1).
    \item Apply the MinT formula to generate reconciled forecasts
\end{enumerate}

Our approach proceeds in a similar way, and can broadly be set out as follows:

\subsection{Revised Approach}

\begin{enumerate}
    \item Collect or separately estimate a set (or several sets) of \emph{exogenous} forecasts any subset of the series in the collection.
    \item Estimate a \emph{baseline} model for the base time series, resulting in in a forecast mean and variance / covariance matrix for these variables.
    \item Using (2), \emph{dis-aggregate} the exogenous forecasts for any aggregates to produce a collection of forecasts for the base time series
    \item Update the distribution in (2) with the relevant dis-aggregated forecasts via a forecast combination regression model.
\end{enumerate}

The remainder of this section describes at a high level how and why we proceed differently to the consensus approach, the next section describes the algorithum in detail.

\subsection{Exogenous Forecasts}

In stage one, we proceed in a similar way to the recent reconciliation literature, in that we collect from external sources and/or independently generate a set of forecasts for the time series within the system. The MINT approach generally considers the situation where one exogenous forecast is provided per series in the collection. Our approach is more flexible, in that we can deal with situations where several forecasts are made for some subset of time series, and no forecasts are available for some other series of interest. 

In general terms, an advantage of a Bayesian approach to forecasting is that it allows sensible forecasts to be made on the basis of very little or no data, using prior experience to form initial views which are then updated given actual experience. In contrast to most of the literature, this flexibility allows us to use out of sample data to revise initial estimates of the forecast combining weights in our revised third step. In this way we place the baseline model, which in general will use a relatively 'conservative' set of hyper-parameters, believed to describe the long run evolution of the system, on a level playing field with base forecasts which are often (certainly in the reconciliation literature) generated by optimising one step ahead predictive ability. In practice we proceed by setting initial priors on regression weights in step (4) which assume that the exogenous forecasts add no value relative to the baseline model. 

Additionally, in this paper we make the process dynamic, allowing for time variation in patterns of data variance / covariance, forecast accuracy and correlation between base forecasts.

\subsection{Baseline Model}

In step 2 we estimate a dynamic multivariate model for the entire system. We regard this as a 'prior' - it should in general terms describe the basic long run time series features of the system, and importantly, result in reasonable estimates of the variance / covariance matrix of the data. To this end, taking in to account the high dimensional nature of the data, we use a factor based approach to estimate this matrix, resulting in dramatic savings both in computational effort and required storage capacity, along with highly desirable shrinkage and sparsity in the resulting covariance matrix.

The essence of the factor based approach to covariance estimation is that relationships between large (or very large) numbers of time series can adequately be described by a relatively small number of common factors $\mathbf{x}_t$. This simple idea the has dramatic mathematical and statistical consequences in terms of minimising the impact of the dimensionality issue we describe above. This can be seen most clearly in terms of the factor model based expression for the data covariance matrix. If we have $n_x$ factors, and $n_a$ base time series, with $n_x<<n_b$ then the data variance / covariance matrix at time $t$ is written as :

\begin{align*}
    Var(\mathbf{b}_t) = \boldsymbol{\Delta}_t \boldsymbol{\Sigma}_t \boldsymbol{\Delta}_t' + \mathbf{D}_t
\end{align*}

Where $\boldsymbol{\Delta}_t$ is a $n_b \times n_x$ loadings matrix, $\boldsymbol{\Sigma}_t$ is the $n_x \times n_x$ covariance matrix of the factors and $\mathbf{D}_t$ is a $n_b$ diagonal matrix of series specific variances. The number of terms in the factor representation of the variance / covariance matrix in this setting reduces to $(n_x \times n_b) + (n_x \times n_x) + n_b$ as opposed to $(n_b \times n_b)$ terms in the absence of factor structure. For a multiline retailer for example, once the factor model is specified, only an additional $(n_x + 1)$ terms are required for each additional SKU, so the computational burden grows linearly with $n_b$. As we describe below, given certina assumptions, we can also  express the reconciled variance / covariance matrix in factor form, so computational benefits extend to the reconciled forecasts too.

In the context of a hierarchy with summation matrix $S$, we note that the covariance matrix of the entire hierarchy can be written as:

\begin{align*}
    Var(\mathbf{y}_t) = \mathbf{S}\boldsymbol{\Delta}_t \boldsymbol{\Sigma}_t \boldsymbol{\Delta}_t'\mathbf{S}' + \mathbf{S}\mathbf{D}\mathbf{S}'
\end{align*}

so that whilst the common component of the covariance matrix retains its factor structure, with loadings given by $\mathbf{S}\boldsymbol{\Delta}_t$, the specific variance matrix $\mathbf{S}\mathbf{D}\mathbf{S}'$ is no longer diagonal. Therefore even if base series are independent (the loadings matrix $\Delta$ is a matrix of zeros), the series specific covariance matrix $SDS'$ adopts a particular non diagonal form driven by the aggregation constraints. 

In our previous work \cite{hollyman2022hierarchies} we utilised \emph{latent} factors, where both the times series of the factors, and the factor loadings were extracted from the data. We associated the latent factors with features of the hierarchy by using prior information to restrict the loadings of the base series to relevant factors. This is a 2 stage process, and a Gibbs sampling procedure was used to extract the factors. In large hierarchies, sampling based approaches are impractical and a different approach is required. In this paper we define the factors as being equal to selected hierarchical aggregates, making them readily by available directly from the data. As \cite{DAgostino2016-bx} point out, latent factors extracted via principal components become equal to cross sectional averages when loadings are assumed to be equal. We then estimate loadings given the subset of aggregates defined as factors, and allow these loadings to change over time. In contrast to our previous work, where we extracted 'static' factors based on residuals from a multivariate model, here the factors are 'dynamic' factors, in the sense that we generate explicit forecasts for them, which in turn drive the forecasts for the base time series. To do this we utilise the Multi-Regression Dynamic Linear Model (MRDLM) structure of (\cite{Queen1993-rn}, \cite{Queen2008-ff}, \cite{Anacleto2012-xy}, \cite{Queen_2008-nx}) further developed in \cite{Zhao2015-mk} and \cite{Zhao2016-ts}. This structure, based on the Dynamic Linear Model (\cite{West1999-am} and references therein) allows for a univariate series to be modelled with the usual structural time series components (level, trend, seasonality etc) along with sets of \emph{contemporaneously} observed 'parental' predictors. Originally developed to encode networks of causality amongst collections of variables, this structure naturally accommodates our factor model setting, with the subset of factors becoming a single parental set for all 'base' series. \cite{Lavine2022-io} adopt just this a approach.

The MRDLM strucutre produces at each time $t$ a filtered mean and covariance matrix for the base time series, along with $h$ step ahead forecasts thereof. 
Our research, along with the empirical results set out below, suggests that MRDLM models do an excellent job 'out of the box' in capturing the dynamics of large hierarchical systems, and do so in a very computationally efficient manner as the DLMs for the base level series can be updated in parallel. For those unfamiliar with DLMs, it is worth emphasising that these models are updated \emph{sequentially}, with the model being updates once after each observation is recorded, so that no optimisation or re-optimisation of models is required. 

The MRDLM structure becomes a baseline, and we look to update this with incremental exogenous forecasts, in a manner familiar from the recent forecast reconciliation literature discussed above. We find that, as expected given the extensive literature on forecast combination, doing so improves the accuracy forecasts produced.

\subsection{Dis-aggregation}

In order to this, we proceed in two steps. Additional forecasts of base time series can be used directly in the forecast combination regressions we describe below. Where additional forecasts are provided in the form of aggregates, we first \emph{dis-aggregate} these to generate the 'top down' forecasts for each base time series corresponding to each relevant aggregate. The 'top down' forecast is generated based on the estimated variance / covariance matrix calculated in step 2.

\subsection{Combination}

Once dis-aggregated forecasts have been generated, we run (dynamic) forecast combination regressions with the objective of updating the baseline forecasts with the incremental information contained in the (dis-aggregated) exogenous forecasts. These regressions takes account of time varying patterns of forecast accuracy, variance and covariance of these potential predictors. This second step requires some simplifying assumptions, but given these can also completed without the need for simulation or sampling. 

In order to further reduce the dimensionality of the problem, for each base time series $b_i$, we consider as predictors in these regressions only the dis-aggregate forecasts derived from an aggregate of which $b_i$ is a component. This generates a very significant reduction in the number of parameters needing to be estimated. In a hierarchical setting, the matrix of regression weights is $n_b \times k$ where k is the number of hierarchical levels (assuming that base time series is a component of precisely one aggregate in each level). In contrast, the MINT weight matrix is of significantly larger dimension - $n_b \times n$.

A further refinement is available if we hypothesise that the regression weights applied to the forecasts derived from each level of the hierarchy are likely so be similar across the base time series. As these regressions take place at the nosiest level of the hierarchy, it is beneficial to pool information to improve statistical estimation of the regression weights. We do this via the hierarchical dynamic regression model described in \cite{Gamerman1993-xw}. 

\subsection{Sub - hierarchies}

We incorporate one further refinement of the process in this paper to aid process scalability. Both the MINT reconciliation approach, and our alternative procedure  require at certain points the inversion of (at least) an $n_b \times n_b$ covariance matrix. This becomes a very significant computational bottleneck for large hierarchies. In order to proceed we break the reconciliation approach down in to stages. An iterative approach to reconciliation was suggested in \cite{Athanasopoulos2017-nw} in the context of cross-temporal hierarchies, and the approach was further developed in \cite{Di_Fonzo2021-tf}. We adopt a similar principal to reconcile a (very) large cross sectional hierarchy, by doing so in two steps.  We take advantage of the insight that a large hierarchy can be broken down in to constituent parts, and then that the reconciled distribution of forecasts in each hierarchy are completely summarised in terms of the means and covariances of the base series.

Consider for example the small, three layer hierarchy described in \cite{Petropolous2024}:

\begin{figure}[!ht]
\begin{center}
\begin{tikzpicture}[level distance=1.7cm,
level 1/.style={sibling distance=5.5cm},
level 2/.style={sibling distance=1.4cm}, 
pool/.style={minimum size=12mm}]
\tikzstyle{every node}=[circle,draw]
    \node[pool] (C) {T}
    child {
        node[pool] {A}
        child { node[pool] {A$_1$} }
        child { node[pool] {A$_2$} }
        child { node[pool] {A$_3$} }
        child { node[pool] {A$_4$} }
    }
    child {
        node[pool] {B}
        child { node[pool] {B$_1$} }
        child { node[pool] {B$_2$} }
        child { node[pool] {B$_3$} }
    }
    child {
        node[pool] {C}
        child { node[pool] {C$_1$} }
        child { node[pool] {C$_2$} }
        child { node[pool] {C$_3$} }
    };
\end{tikzpicture}
\end{center}
\caption{A small, three-level hierarchy.}
\label{fig:sh}
\end{figure}
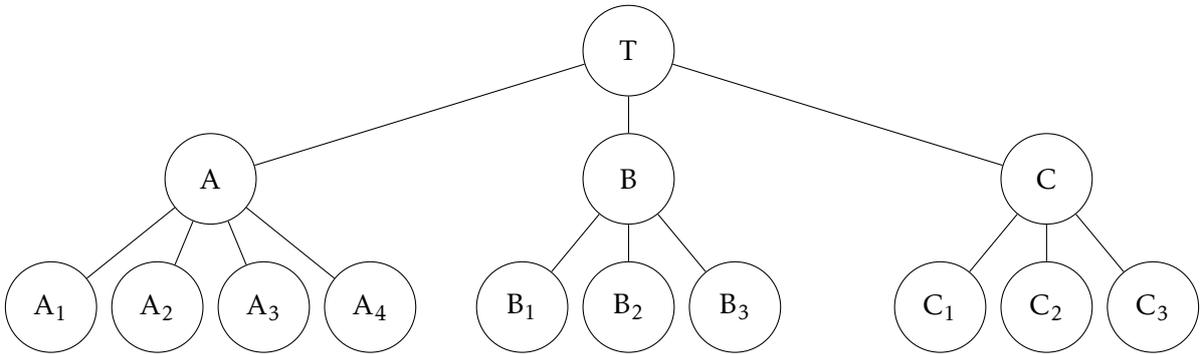

Given a set of exogenous forecasts for each time series, we proceed to reconcile this hierarchy as follows. Firstly we estimate the MRDLM model for the base series $[A_1...C_3]$, potentially using the aggregates $[A,B,C]$ (or perhaps just the single series $T$) as factors. 

\begin{figure}[H]
\begin{center}
\begin{tikzpicture}[level distance=1.7cm,
level 1/.style={sibling distance=1.4cm},
pool/.style={minimum size=12mm}]
\tikzstyle{every node}=[circle,draw]
    \node[pool] {T}
        child { node[pool] {A} }
        child { node[pool] {B} }
        child { node[pool] {C} };
\end{tikzpicture}
\end{center}
\caption{Upper sub-hierarchy.}
\label{fig:tsh}
\end{figure}
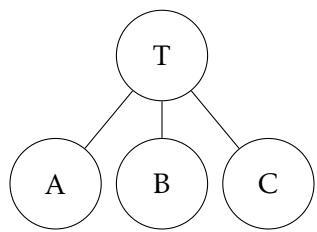

Using the MRDLM variance / covariance matrix, We then  dis-aggregate the exogenous aggregate forecasts (i.e. those of series $[T,A,B,C]$) resulting in three forecasts for each of the base series (one derived from each hierarchical level). We now run a forecast combination regression for the top sub - hierarchy (composed of the series $[T,A,B,C]$ as shown in figure \ref{fig:tsh}), using the series $[A,B,C]$ as the base time series for this sub-hierarchy. We derive the regressors by applying the appropriate $\mathbf{C}$ matrices for each \emph{lower} sub-hierarchy to the two sets of dis-aggregated forecasts (derived from $T$ and from $[A,B,C]$).

This results in a forecast mean and variance /covariance matrix for series $[A,B,C]$, which completely summarises the optimal reconciled forecasts for the upper sub-hierarchy, given the exogenous forecasts at these levels.

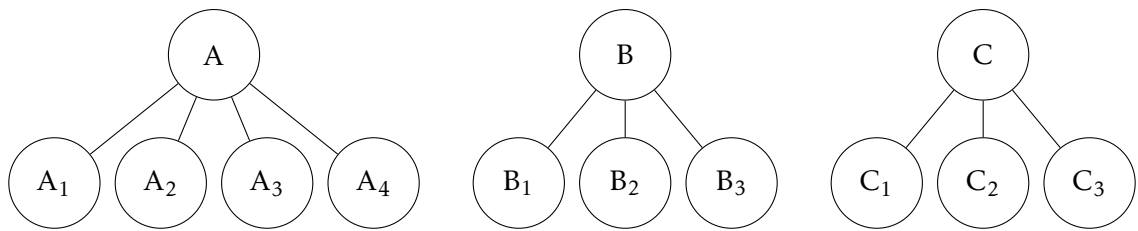
\begin{figure}[!ht]
\begin{center}
\begin{tikzpicture}[level distance=1.7cm,
level 1/.style={sibling distance=1.4cm},
pool/.style={minimum size=12mm}]
\tikzstyle{every node}=[circle,draw]
    \node[pool] {A}
        child { node[pool] {A$_1$} }
        child { node[pool] {A$_2$} }
        child { node[pool] {A$_3$} }
        child { node[pool] {A$_4$} };
\end{tikzpicture}
	\hspace{5mm}
\begin{tikzpicture}[level distance=1.7cm,
level 1/.style={sibling distance=1.4cm},
pool/.style={minimum size=12mm}]
\tikzstyle{every node}=[circle,draw]
    \node[pool] {B}
        child { node[pool] {B$_1$} }
        child { node[pool] {B$_2$} }
        child { node[pool] {B$_3$} };
\end{tikzpicture}
	\hspace{5mm}
\begin{tikzpicture}[level distance=1.7cm,
level 1/.style={sibling distance=1.4cm},
pool/.style={minimum size=12mm}]
\tikzstyle{every node}=[circle,draw]
    \node[pool] {C}
        child { node[pool] {C$_1$} }
        child { node[pool] {C$_2$} }
        child { node[pool] {C$_3$} };
\end{tikzpicture}
\end{center}
\caption{The three separate, lower sub-hierarchies}
\label{fig:sh1}
\end{figure}

We now treat these reconciled forecasts as a exogenous forecasts for the middle level, and disaggregate them to base series $[A_1...C_3]$, where we run three separate reconciliation regressions, one for each sub-hierarchy shown in figure \ref{fig:sh1}, where the regressors are the now dis-aggregated reconciled forecasts from level 2, and the original exogenous forecasts for level 3. The bottom level hierarchies are considered independent given the top level, and can therefore be reconciled in parallel. This enables significant scaling of the HF approach to (very) large hierarchies. 

In this example it is natural to fit the forecasts derived from each level in to the corresponding top and bottom sub-hierarchical reconciliation regressions, however we note that forecasts from any given level can potentially be included arbitrarily in either sub-hierarchy, separating the role of forecast information and forecast constraints in the reconciliation process. This potentially this allows for the regressions at the noisier bottom level of the hierarchy to be estimated using a smaller number of regressors, although we do not explore this option further in this paper.

The resulting 2 - stage forecasts incorporate all of the forecast information from all of the hierarchical levels, and (contingent on the structure of the hierarchy - a topic we return to below) the reconciled bottom level series are optimal given the constraints applying to each bottom level hierarchy. They may \emph{not} in theory however optimally satisfy the constraints on the forecast regression in the upper sub-hierarchy. In our research, any further improvements in accuracy were well within the bounds of uncertainty, and running further iterations of the procedure generated no real practical benefit in terms of forecast accuracy, so we do not so here.

We now move on to describe the various modelling steps in greater mathematical detail.

\section{The model}

\subsection{Multiple Regression Dynamic Linear Models}

The MRDLM model can be described as follows.  Firstly, the system is driven by a small number $n_x$ of factors, where $n_x$ is small in relation to the number of base series $n_b$ and the total number of aggregates $n_a$. We define the factors, which we denote as $\mathbf{x}_t$ are observable as subset, chosen by the analyst, of the aggregates $\mathbf{a}_t$. The factors are then modelled as a General Multivariate DLM \cite{West1999-am}, \cite{Triantafyllopoulos2007-hr}:

\begin{align*}
   \mathbf{x}_t & = F_{x,t}'\theta_{x,t} + e_{x,t} & e_{x,t} \sim N(0,V_{x,t})\\
   \theta_{x,t} & = G_{x,t} \theta_{x,t-1} + w_{x,t} & w_{x,t} \sim N(0,W_{x,t})
\end{align*}

Where $x_t$ is the $n_x$ vector of factors at time t, $\theta_{x,t}$ is the $q_x$ parameter / state vector at time $t$, $F_{x,t}$ is an $q_x \times n_x$ matrix of regressors (usually level / trend and seasonal time series components) and $G_{x,t}$ is the $q_x \times q_x$ state evolution variance at time $t$. The observation and state innovation variances are $V_{x,t}$ and $W_{x,t}$ respectively, with  $W_{x,t}$ being specified via discount or forgetting factors (\cite{West1999-am}).  The General Multivariate DLM has the advantage of being more flexible than Matrix Normal Multivariate Structure (\cite{Triantafyllopoulos2007-hr}, \cite{Triant-phd}) used the the hierarchical forecasting context by \cite{hollyman2022hierarchies} and can be adapted to incorporate both random and deterministic changes in variance via variance laws (\cite{West1999-am}). This flexibility comes at a cost in that there is no neat conjugate analysis available in the usual scenario of unknown $V_{x,t}$. One solution to this issue is to adopt a simulation based approach, which would be feasible given the relatively small number of factors, but instead in this paper we utilise the closed form variance approximation developed for such models in \cite{Triantafyllopoulos2007-hr} which is computationally \emph{very} efficient here given the small size of the factor covariance matrix. For increased robustness in our results we allow for a degree of random variation in variance estimates via the variance discounting mechanism set out in \cite{West1999-am} and  \cite{Prado2021-xl}.

We now turn to the base time series. These are modelled as follows using $n_b$ separate univariate DLMS:

\begin{align*}
   b_{i,t} & = F_{i,t}\theta_{i,t} + e_{i,t} & e_{i,t} \sim N(0,v_{i,t})\\
   \theta_{i,t} & = G_{i,t} \theta_{i,t-1} + w_{i,t} & w_{i,t} \sim N(0,W_{i,t})
\end{align*}

Where here, the state vector $\theta_{i,t}$ for each separate model incorporates both time series components and loadings on the  contemporaneous factors $\mathbf{x}_t$ (or an appropriate subset thereof) now embedded in each $F_{i,t}$. The DLM structures and discount rates adopted for the factor and base series will in general be different. Cross sectional aggregation generally smooths out noise so that aggregate series will generally have a higher signal to noise ratio than the base time series. In general therefore we can include a higher number of time series components in the aggregate models, allowing for trends, and potentially auto-regressive components and exogenous regressors (although we do not pursue these options here). Additionally, the aggregate models can be specified using more aggressive discount factors to enable changes in time series components to be picked up more quickly. DLMs for the noisier base series will comprise basic level and simple seasonality components, along with a small number of regression components to pick up effects from the factors, and capture cross sectional covariance. They will generally require less aggressive discount or forgetting factors to allow for a longer history of temporal information to be retained in the state vectors.

Both sets of models are fitted using standard Kalman filter recursions set out for example in (\cite{West1999-am}, \cite{Prado2021-xl}), although in practice we adopt singular value decomposition (SVD) based calculations for numerical stability. Calculations for the variance approximation of \cite{Triantafyllopoulos2007-hr} are set out in that paper, and also fit neatly in to the SVD based Kalman recursions we use. Calculation of the individual base level DLMs can proceed in parallel once the factor DLM has been updated. Updating proceeds sequentially and calculations are typically rather fast with modern hardware and software (we utilise the Python JAX library \cite{jax2018github}) even for large or very large systems.

Once the time $t$ Kalman filter updates of the factor model and each univariate model have been completed, the full data covariance structure is derived using a separate process, originally set out in \cite{Queen1993-rn}, in then in some detail in \cite{Zhao2016-ts}. Both authors describe a general graphical model case where a series may have several generations of parental series, and the matrix is built up in an iterative fashion starting with parental nodes and progressing series by series. In our application, the calculations simplify quite significantly as we are concerned only with one set of parental relationships, namely those between the base series and a sub-set of aggregates which double up as factors, so that given the factor covariance matrix the covariance terms for the base series can be computed in a vectorised manner as described in \cite{Lavine2022-io} for Generalised DLMs. For the MRDLM model we produce $h$ step ahead forecast means for the base series. The forecast covariances between the series are retained in factor form, so we effectively produce a $h$ step ahead $(n_x \times n_x)$ factor covariance matrix, an $(n_b \times n_x)$ loadings matrix and a $n_b$ vector of specific series variances for each time horizon $h$ so that the $h$ step ahead prior forecasts can be written as:

\begin{align}
    \label{h-step ahead prior}
    \bar{\mathbf{b}}_{t,t+h} &\sim  \left[ \bar{\mathbf{f}}_{t,t+h}, \boldsymbol{\Delta_t} \boldsymbol{\Sigma}_{x,t,t+h} \boldsymbol{\Delta}'_t  + \mathbf{D}_{t,t+h}\right]
\end{align}

With $\mathbf{D}_{t,t+h}$ denoting a diagonal matrix of series specific variance forecasts.

\subsection{Incorporating additional exogenous forecasts - Dis-aggregation and Combination}

At any time $t$ the MRDLM produces a set of $h$ step ahead forecasts and an associated covariance matrix. In the reconciliation setting popular in the recent literature, we have available a set of exogenous forecasts, in addition to the MRDLM forecasts and variance / covaraince matrix described above. The approach adopted in this paper is to treat the MRDLM as a 'prior' which describes the basic long run behaviour of the system, accounting for basic time series features, which is then updated with the additional exogenous forecasts. Usually in the existing literature one exogenous forecast is provided for each of the $(n_a+n_b)$ series in the collection. Our alternative is significantly more flexible in that it allows for one or more sets of forecasts for any subset of the series in the hierarchy, although for simplicity of exposition we follow convention for the rest of the paper and utilise one set of incremental forecasts for each of the series in the hierarchy.

We proceed in three steps. Firstly, we derive revised forecast distributions for each series in $\mathbf{y}_t$ given forecast information $\hat{\mathbf{y}}_t$. Secondly we dis-aggregate this revised information to the base series level, estimating a revised view of each given the exogenous forecast information. Finally, for each base series, we estimate a combination regression with the dis-aggregated forecasts from each of its hierarchical 'parents' as regressors. The weights estimated in these regressions take account of the linear constraints implied by the hierarchy via the $\mathbf{S}$ matrix, i.e. they are chosen to optimise the accuracy of the reconciled forecasts of the aggregate time series, as well as the those of the base time series.

The procedure for updating an existing forecast distribution for additional exogenous forecasts is clearly set out in \cite{West1999-am}, and the following exposition closely follows the approach set out therein. We assume the availability of $h$ step ahead exogenous forecasts, aligning with the prior forecasts in \ref{h-step ahead prior}, although we suppress the $t+h$ subscripts in what follows.

In our case the MRDLM model provides the required multivariate prior distributions from (\ref{h-step ahead prior}) which we denote in this section by $\bar{\mathbf{b}}$:
\begin{align*}
    \bar{\mathbf{b}}_t \sim [\bar{\mathbf{f}}_{bt},\bar{\mathbf{Q}}_{bt}]\\
\end{align*}
Consider the case where an exogenous forecast $\hat{a}_{it}$ is available for an aggregate time series $i$ so that the aggregate series is given by $a_{it} = \mathbf{c}_i\mathbf{b}_t$, where $\mathbf{c}_i$ is a known $n_b$ vector. Relating this back to the reconciliation approaches of recent years $\mathbf{c}_i$ will correspond to the $i$th row of the aggregation matrix $\mathbf(S)$, with $i \leq n_a$. Before observing the value of the exogenous forecast for $a_{it}$ the (singular) joint distribution of $[\bar{\mathbf{b}}_t,\bar{{a}}_{it}]'$ is: 
\begin{align}
    \left[\begin{matrix}\mathbf{\bar{b}}_{t} \\ 
    \bar{a}_{it}\end{matrix}\right] 
    \sim
    \left[\left(\begin{matrix}\bar{\mathbf{f}}_{bt} \\ \mathbf{c_i}\bar{\mathbf{f}}_{bt}\end{matrix}\right) ,\left(\begin{matrix}\bar{\mathbf{Q}}_{bt} & \bar{\mathbf{Q}}_{bt}\mathbf{c}_i \\ \mathbf{c}_i\bar{\mathbf{Q}}_{bt} & \mathbf{c}_i \bar{\mathbf{Q}}_{bt} \mathbf{c}_i\end{matrix}\right)\right]
\end{align}
We wish to update this distribution given exogenous forecast information $\hat{a}_{it}$. The procedure adopted in \cite{West1999-am} does this in two steps, although the first step described by these authors effectively drops away in the procedure we adopt in this paper, as we now set out.

\subsubsection{Dis-aggregation}

Firstly we estimate the posterior distribution of $a_{it}|\hat{a}_{it}$, i.e the forecast distribution of the series in question $a_{it}$ given new forecast information $\hat{a}_{it}$. We adopt the model discussed by these authors, based on earlier work in \cite{West1988-rp}, \cite{West1992-mg}, \cite{West1992-ml} so that given external forecast information $\hat{a}_{i,t} \sim [\hat{f}_{it},\hat{q}_{it}]$ the revised forecast of $a_{it}$, which we denote by $\hat{a}^*_{it}$ has mean and variance:
\begin{align*}
    \hat{f}^*_{it} &= \mathbf{c}_i\bar{\mathbf{f}}_{bt}+\rho\left(\hat{f}_{it} - \mathbf{c}_i\bar{\mathbf{f}}_{bt}\right)
    &
    \hat{{q}}^*_{it}  &= 
    \left(1-\rho^{2}\right) \mathbf{c}_i \bar{\mathbf{Q}}_{bt} \mathbf{c}_i+\rho^2 \hat{q}_{it}\\
\end{align*}
The quantity $\rho$ is a measure of the quality of the forecast information provided. If $\rho$ is zero, the additional forecasts are ignored, and the forecast distribution of $y_t$ remains unchanged post receipt of the external forecast information. On the other hand, when $\rho = 1$ the forecast information is adopted in full, and the prior forecast is disregarded. In a previous paper \cite{hollyman2022hierarchies} we estimated $\rho$ from the data, but here we take $\rho = 1$ and adopt the forecast information $\hat{a}_{i,t} \sim [\hat{f}_{i,t},\hat{q}_{i,t}]$ in full and at face value. The implication is that we make no attempt to calibrate the base forecasts at the dis-aggregation stage, leaving this to a final combination regression step, where the prior forecasts of the base time series are updated as we describe below.

The procedure described above focuses on the case of a single univariate forecast. The analogous procedure for vectors of forecasts is set out in \cite{West1999-am}, but these require estimates of covariance for the exogenous forecasts. It would be possible to estimate the forecast error covariance matrix with a shrinkage procedure similar to tat adopted in MinT type approaches (or otherwise) but we do not pursue that option here as this becomes increasingly intractable as the size of the hierarchy increases, and there are additional computational  benefits of not doing so, whihc we outline below. We account for the covariance of \emph{dis-aggregated} forecasts in the combination regressions we describe below.  

\cite{West1999-am} now describe how to  dis-aggregate the revised forecast of $a_{it}$ to derive the updated forecast distribution of $\mathbf{b}_{t}$ given $\hat{a}_{it}$. As we make no adjustment to the exogenous forecast $\hat{a}_{it}$,  we apply this procedure directly to that quantity.  As set out by the original authors, we assume that given $a_{it}$ the base series $\mathbf{b}_t$ are conditionally independent of the new information $\hat{a}_{i,t}$, so $p(\mathbf{b}_t|\bar{a}_{i,t},\hat{a}_{i,t}) = p(\mathbf{b}_t|\bar{a}_{i,t})$. This assumption effectively states that receipt of the new forecast information regarding $a_{it}$ tells us nothing further about how to dis-aggregate estimates of this variable to the base level series - the dis-aggregation process is does not depend on $\hat{a}_{i,t}$.  Then:

\begin{align}
    \label{diaggregation}
     (\mathbf{b}_t|\hat{a}_{i,t}) &\sim [\bar{\mathbf{f}}_t + \mathbf{{q}}_i(\hat{f}_{i,t} - \bar{f}_{i,t})/\bar{Q}_i, \bar{\mathbf{Q}}_{i,t} -\mathbf{{q}}_{i,t}\mathbf{{q}}_{i,t}^{\prime}(\bar{q}_{it} - \hat{q}_{it})/\hat{q}^{2}_{it}]
\end{align}

With $\bar{\mathbf{q}}_i = {\bar{\mathbf{Q}}}\mathbf{c}_i$ and  $\bar{q}_{it} = \mathbf{c}_i\bar{\mathbf{Q}}_{it}\mathbf{c}_i$.

The above discussion focuses on dis-aggregation of an aggregate time series. The same process applies to any base time series $i$, with a vector $\mathbf{c_i}$ as one row of $I_{n_b}$ i.e. a row of zeros except for a one in position $i$.

The practical implication of (\ref{diaggregation}) is that for any series $y_{it}$, observation of the exogenous forecast $\hat{y}_{it}$ leads to the revision of \emph{each} base series $b_{jt}$, $j=1...n_b$. This is an important point, which we illustrate by example (we drop the time subscripts on the variables below for ease of exposition). 
Consider a simple hierarchy composed of three series, two base ($n_b = 2$) time series A and B, and a total, T, such that T = A + B. The total number of series in this simple hierarchy is $n = n_b+n_a =  3$.

The matrices $\mathbf{C}$ and $\mathbf{S}$ for this hierarchy are as follows, and we assume that we have a prior for $b \sim \left[ \bar{\mathbf{f}}_b, \bar{\mathbf{Q}}_b \right]$ as given below.

\begin{align*}
 \mathbf{C} &=\left[ \begin{matrix} 1 & 1 \end{matrix} \right] & 
 \mathbf{S} &=\left[\begin{matrix}\mathbf{C} \\\mathbf{I}_2 \end{matrix}\ \right]\\\\
 \bar{\mathbf{f}}_b =&\left[ \begin{matrix} 0 \\ 0 \end{matrix} \right] &
 \bar{\mathbf{Q}}_{b} =&\left[ \begin{matrix}  1 & 0.5 \\ 0.5 & 1 \end{matrix} \right] \end{align*}

Assume we have an external forecast for base series $A$ so that $\mathbf{c}_i = \left[ \begin{matrix} 1 & 0 \end{matrix} \right]$. It follows that $\bar{\mathbf{q}}_{A} = \left[ \begin{matrix} 1 & 0.5 \end{matrix} \right]$ and $\bar{q}_{A} = 1$. We now observe the forecast which holds that $A \sim \left[ 0.1 , 0.9 \right]$. The revised forecast distribution of $\mathbf{b}$ from (\ref{diaggregation}) is then: 

\begin{align*}
 \left[ \left[ \begin{matrix} 0.1 & 0.05 \end{matrix} \right] ,\left[ \begin{matrix} 0.9 & 0.45 \\ 0.45 & 0.975\end{matrix} \right]\right]
\end{align*}

Hence when we observe a forecast for base series A, we logically \emph{also} make a small revision to the forecast distribution of B, as A and B are correlated.

The same situation arises when a revision is made to some aggregate series $a_{it}$ - such a revision results in small changes to the forecast distribution all series in $b_t$. 

In this simple example, when the distributions of $\mathbf{b}$ is assumed known, it makes sense that because of the relatively substantial level of correlation between series A and B, incremental information relating to series $A$ results in a revised distribution for $B$. The implication of this is that each of the $n$  forecasts in $\hat{\mathbf{y}}_t$ results in a separate revised distribution for each of the $n_b$ series in $\mathbf{b}_t$. We argue that in real world scenarios where correlation / covariance between the often rather noisy base series needs to be estimated, \emph{and} we have available separate 'bespoke' forecasts available for each series, it makes sense to shrink forecasts for any given base series to zero where there is no hierarchical relationship between $b_i$ $i=1...n_b$ and $y_{j}$ $j = 1 ... n$. Where base series $i$ has $k_i$ hierarchical parents, this means that the reconciled forecasts for each base series $y_i$ are estimated as combinations of $k_i<<n$ forecasts, offering a dramatic reduction in the dimensionality of the reconciliation problem.

\subsubsection{Forecast Combination}

From above, we have for each base series $b_{it}$ a set of $k_i$ forecasts, one derived from each node in of which series $i$ is a component. For the remainder of the paper we assume that we have k forecasts for each base series, typically one for each level in the hierarchy. The final step in the reconciliation process is to estimate a combination regression with these $(n_b \times k)$ forecasts as regressors. The regression coefficients represent the weights allocated to the forecast derived from each level, and serve to calibrate and combine each set of forecasts. We estimate these regressions using dynamic linear models, to allow for potential time varying patterns of forecast accuracy, and inter series correlation. Differently to the existing literature on forecast reconciliation we estimate forecast combination regressions based on out of sample prior and exogenous forecasts. In this paper we do so by estimating the observed outcomes at each time t as a function of the one step ahead prior and exogenous forecasts made at time $t-1$. It is a relatively simple matter to perform separate (perhaps hierarchically structured) regressions for each time horizon $h$, although we do not do so here. 

In general, the forecast combining regression for an individual, base series can also be written as (yet another) DLM. In the hierarchical setting it is important to choose the weights in $\theta$ accounting for the aggregation constraints, so the combination regressions can not be estimated equation by equation. We therefore proceed to write the regression as a multivariate DLM in the base series $\mathbf{b}_t$:

\begin{align}
\label{mv-comb-regression} \mathbf{b}_{t}&=\mathbf{F}'_{t} \boldsymbol{\theta}_{t}+\boldsymbol{\varepsilon}_{t} &
\quad \boldsymbol{\varepsilon}_{t} &\sim N\left(0, \mathbf{V}_{t}\right) \\ \boldsymbol{\theta}_{i,t}&=\boldsymbol{\theta}_{t-1}+\boldsymbol{\omega_{t}} &\quad \boldsymbol{\omega}_{t} &\sim N\left(0, \mathbf{W}_{t})\right.
\end{align}

Where here, $\mathbf{F}'_t$ is an $n_b \times (n_b \times k)$ block diagonal matrix with the regressors for each base series as diagonal blocks, and $\boldsymbol{\theta}_t$ is an $(n_b \times k)$ vector comprising the regression weights.

In order to estimate $\boldsymbol{\theta}_t$ accounting for the aggregates we pre-multiply by $\mathbf{S}$ and write:

\begin{align}
\label{hier-mv-comb-regression} \mathbf{S}\mathbf{b}_{t}&=\mathbf{S}\mathbf{F}'_{t} \boldsymbol{\theta}_{t}+\mathbf{S}\boldsymbol{\varepsilon}_{t}  & \quad  \mathbf{S} \boldsymbol{\varepsilon_{t}} &\sim N\left(0, \mathbf{S}\mathbf{V}_{t}\mathbf{S}'\right) \\  \boldsymbol{\theta}_{t}&=\boldsymbol{\theta}_{t-1}+\boldsymbol{\omega_{t}} &\quad \boldsymbol{\omega}_{t} &\sim N\left(0, \mathbf{W}_{t})\right.
\end{align}

Which is a DLM in $\mathbf{S}\mathbf{b}_{t}$ and $\mathbf{S}\mathbf{F}'_{t}$ with the same parameter vector $\theta$. Closed form estimation of such multivariate regressions involves the use a number of approximations. Firstly we proceed with estimation assuming that the error variance / covariance matrix is known, and equal to $\hat{\mathbf{Q}}_t$, that estimated from the MRDLM model set out above.  Such assumptions, using pre-estimated covariance matrices has a long history in Bayesian time series analysis; used for example in Vector Auto Regressions utilising the Minnesota prior (\cite{Doan1984-ju}, \cite{Litterman1986-lq}). Once the estimate of the weight coefficients is made, we update the estimate of $\hat{\mathbf{Q}}_t$ to $\hat{\mathbf{Q}}^*_t$ using a linear bayes approximation (\cite{Goldstein2007-it}).

Standard dynamic regression assume the regressors in $F_t$ are known. In the context discussed here, the regressors are the exogenous base forecasts, propagated to the bottom layer via a noisy process, so even when $\hat{\mathbf{y}}_{it}$ are supplied as point forecasts, we treat them as random, and adopt an approximation based on conditional expectation similar to that set out in \cite{Lavine2022-io} for exponential family DLMs (see also \cite{West1999-am} page 279). These approximations lead to incremental benefits in terms of the expression for the reconciled forecast variance / covariance matrix of the base series, because the prior to posterior update of the  variance /covariance matrix of the reconciled series is then expressed as a $n_b \times n_b$ diagonal matrix, which can be stored as an $n_b$ vector and readily applied to $\hat{\mathbf{Q}}_t$ which is already in quadratic or 'factor model' form. This approach is consistent with the idea of univariate exogenous forecasts, which may be informative regarding the location and variance of the respective series, but might be expected to be less so regarding covariances between them. We describe this part of the the estimation of this regression in detail in Appendix F.

We introduce one further element in to the estimation of the combination weights. The base level series are inherently the noisiest and most difficult to estimate and forecast, yet we must estimate combination weights at this level. We therefore employ a hierarchical prior, which serves to increase estimation accuracy by pooling information across series. Such priors, in the context of dynamic linear models were formalised by \cite{Gamerman1993-xw}. We utilise this hierarchical prior at the sub-hierarchy level, leaving the top hierarchy to be estimated as above.

In this context we adopt a two level hierarchy, where the observations at base level are described by an observation equation:

\begin{align}
    \mathbf{b}_{t}&=\mathbf{F}_{bt} \boldsymbol{\theta}_{bt}+\mathbf{v}_{bt} &
    \quad \mathbf{v}_{bt} &\sim N\left(0, \mathbf{V}_{bt}\right)
\end{align}

The vector of regression parameters for each base series $\boldsymbol{\theta}_{bt}$ is related to a shared, hierarchy wide set of parameters $\boldsymbol{\theta}_{ht}$ via a structural equation:

\begin{align}
    \boldsymbol{\theta}_{bt}&=\mathbf{F}_{ht} \boldsymbol{\theta}_{ht}+v_{ht} &\quad v_{t} &\sim N\left(0, \mathbf{V}_{ht})\right.
\end{align}

The hyper parameter $\mathbf{V}_{ht}$ controls the degree to which series level combination weights differ from their hierarchy wide estimates. These shared parameters then evolve over time as:

\begin{align}
\boldsymbol{\theta}_{ht}&=\boldsymbol{\theta}_{h,t-1}+\boldsymbol{\omega_{t}} &\quad \boldsymbol{\omega}_{t} &\sim N\left(0, \mathbf{W}_{t})\right.
\end{align}.

In the above, $\mathbf{F}_{bt}$ are the per base series regression parameters, comparable to those in (\ref{mv-comb-regression}) and $\mathbf{F}_{ht}$ is s known matrix mapping the shared, hierarchy wide regression weights $\boldsymbol{\theta}_{ht}$ to the series level weights $\boldsymbol{\theta}_{bt}$. Estimation of the parameters of this model is possible in closed from, assuming knowledge of the series covariance matrix $\mathbf{V}_{bt}$, and we make the same assumption as described above, i.e. that $\mathbf{V}_{bt}$ is known and equal to the MRDLM covariance matrix $\bar{\mathbf{Q}}_{bt}$. Further estimation details are set out in Appendix G.

\section{Illustration - Australian Domestic Tourism Dataset}

\subsection{Stand-alone MR-DLM Model}

We begin by illustrating the effectiveness of the stand alone MRDLM algorithm on the Australian Domestic Tourism data set \cite{Athanasopoulos2009-xs}. We illustrate our results using three sets of discount rate hyper-parameters selected from conventional ranges, broadly corresponding to 'fast' 'medium' and 'slow' levels of adaption to new information. Because the top level series we select as factors are aggregates, we expect them to contain more information, and less noise than the base series of which they are comprised, so we use more aggressive discount rates for these series. We also select time series models with a trend component for the factor series, whereas the 'base' series are modelled with simple level and seasonality components. We follow convention in allowing time series seasonal components to evolve more slowly than level and trend. Additionally we specify time varying variances using the variance learning model described in \cite{West1999-am} with a fixed discount parameter of .99. Results are set out in full in Appendix A, and summarised in figure \ref{fig:MR-DLM Tourism Forecast Accucary}.

The set up of the forecasting experiment echos that in \cite{Wickramasuriya2019-av}, we use the first 96 observations to train the MRDLM model, and produce a first set of 12 month ahead quasi out-of-sample forecasts from that point on. We then perform sequential monthly updates to the model, and generate new sets of 12 month ahead forecast after each monthly update, in contrast to the MINT approach which uses a rolling 96 month window and re-fits models at each point in time. Initial state priors are based on estimates of means and variances of level and seasonality from the in-sample period, with state variances set to the estimated variance / 10. As is normal for DLM models, the influence of the initial priors declines quite quickly during the initial estimation period, and results are not sensitive to these sssumptions.

\begin{table}[h]
\small
\centering
\caption{Discount rate parameters}
\begin{tabular}{cccccccc}
\hline
\textbf{ } &
\textbf{Factor}&
\textbf{Factor} &
\textbf{Base} &
\textbf{Base} &
\textbf{Base}\\
\textbf{Category} &
\textbf{Level/Trend}&
\textbf{Seasonality} &
\textbf{Level} &
\textbf{Seasonality} &
\textbf{Regression}\\
\hline
Fast & .9 & .95 & .95 & .97 & .97\\
Medium & .95 & .97 & .97 & .99 & .99\\
Slow & .97 & .99 & .99 & .995 & .995\\
\hline
\end{tabular}
\label{tab:h10t1}
\end{table}

To asses point forecast accuracy we calculate RMSE across the forecast period and express this as a multiple of our 'base case' forecasts, which we define to be bottom up aggregates of the base forecasts. To assess probabilistic forecast accuracy, we calculate the log scores of all the methods (note that we only consider coherent sets of forecasts, so log scores are proper in this context \cite{Panagiotelis2021-sm}). We calculate log-scores assuming gaussian disturbances, utilising the step ahead forecast variance / covariance matrices from each method, and display results averaged across all time series within each hierarchical layer. We again express these as multiples of our base case forecasts. We produce forecasts on a monthly basis but report results averaged over forecast quarters.

We compare our MRDLM models to a number of variants of several well known benchmarks from the literature. Firstly we include 'bottom up' forecasts, which are by definition coherent. We include bottom up forecasts with a diagonal covariance matrix ('BU-Diag'), and also the diagonal target shrinkage covariance matrix (\cite{Schafer2005-wi}) used in \cite{Wickramasuriya2019-av} but in this instance applied only to the base level forecasts ('BU-Shrink'). Whilst these forecasts have identical means they differ notably in distributional forecast accuracy under aggregation, where the shrinkage procedure leads to significant improvements in scores.

\begin{figure}[h]
\centering
\includegraphics[width=0.75\textwidth]{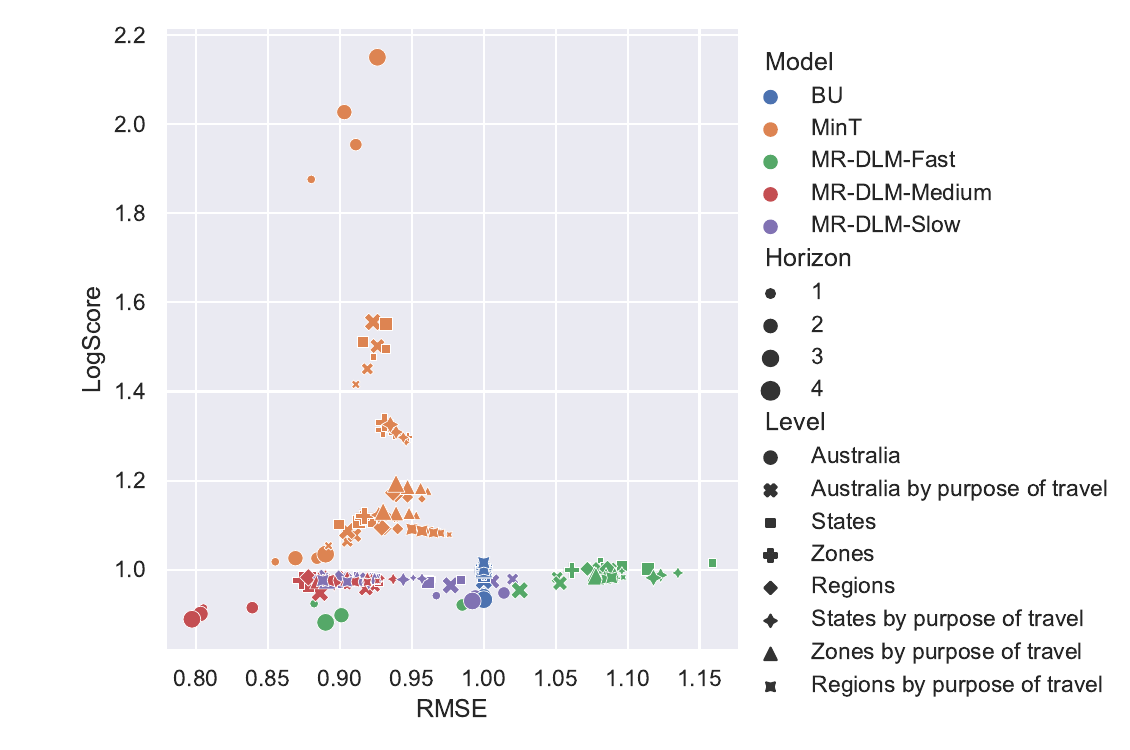}
\caption{We display Log Score (y axis) and RMSE (x axis) relative to the base case of bottom up aggregate forecasts. For full discussion of the results see main text. }
\label{fig:MR-DLM Tourism Forecast Accucary}
\end{figure}

Secondly we explore the well known MINT optimal forecast reconciliation method of \cite{Wickramasuriya2019-av}, which is known to perform well on  a number of hierarchical forecasting datasets. The MINT approach typically estimates the reconciled forecasts distribution of the base series as a function of the time $t$ estimated forecast error variance / covariance matrix $\mathbf{W}_t$ (\emph{Lemma 1} and \emph{Theorem 1} of \cite{Wickramasuriya2019-av}):

\begin{align}
    && \tilde{\mathbf{b}}_{t+h} &\sim \left[\mathbf{G}\hat{\mathbf{y}}_{t+h}, \boldsymbol{\Sigma}^y_{t+h}\right] &&&
    \mathbf{G} &= (\mathbf{S'W}^{-1}_t\mathbf{S})^{-1}
    \mathbf{S'{W}}^{-1}  &&& \boldsymbol{\Sigma}^y_{t+h} &=  \mathbf{SGW}_t\mathbf{G'S'}&
    \label{Mint}
\end{align}

As these authors state 'application of this result depends on having a reliable estimator
for $\mathbf{W}_h$' ($\mathbf{W}_t$). Unfortunately, and as outlined above, such estimates are often not readily available. We include three flavours of MinT - 'OLS' (setting $\mathbf{W}_t=\mathbf{I}_m$) and 'WLS' (setting $\mathbf{W}_t$ as a diagonal matrix of empirical forecast variances) and a third, shrinkage based approach utilising a shrinkage estimator described in \cite{Schafer2005-wi}. The literature has focused mostly on the point forecast accuracy of the MinT approach, and OLS and WLS variants often approach or exceed the point forecast accuracy of shrinkage based MinT estimates. Our analysis shows that distributional forecast accuracy is more heavily effected by noisy and potentially mis-specified estimates of $\mathbf{W}_t$. In our experiments, MinT-OLS estimates produce wildly inaccurate and essentially meaningless estimates of forecast variance / covariance. As the size of the datasets we target with our algorithm are too large to make sampling based approaches feasible, we use the closed form forecast error variance expressions in (\ref{Mint}). As above, OLS produces reasonable \emph{point} forecast estimates, so that if we were to use the Energy score probabilistic scoring rule (\cite{Gneiting2007-ya}, \cite{Panagiotelis2021-sm}) which uses an approach based on sampling realisations of the point forecasts, the performance of the OLS approach would be likely to improve significantly. \cite{Wickramasuriya2024-qu} conduct a number of experiments on an updated and expanded (in terms of time span) version of the Tourism data, but include a smaller number of series, excluding the noisier bottom level observations, and expand the estimation window form 96 to 120 months. They demonstrate a more respectable performance of the OLS approach in terms of log score probabilistic forecasting. Based on these results however, we caution against using clearly miss-specified heuristic approximations to the forecast error covariance matrix in conjunction with error variance estimates in  (\ref{Mint}).

The MinT-Shrink algorithm generally improves both point and distributional forecast accuracy, although not to the extent described in \cite{Wickramasuriya2024-qu} on the smoother higher level aggregates reported in that paper.

\cite{Panagiotelis2021-sm} and \cite{GIROLIMETTO2023} develop a number of probabilistic approaches based on the MINT algorithm, which are shown to improve distributional forecast accuracy, but require computationally expensive simulation or optimisation procedures, which are not practically applicable to the scale of problem we seek to address in this paper.

The MRDLM based approaches are competitive with all of the existing alternatives.  In terms of point forecast accuracy, the 'medium' speed models perform creditably. The faster models perform well on upper hierarchy levels, but somewhat less so towards the bottom of the hierarchy. In terms of distributional accuracy, each of the MRDLM models improves notably on MinT based variants, it is interesting to note that distributional accuracy is relatively robust to the choice of discount parameter (and hence the speed of state evolution) in the DLM models.

\subsection{Incorporating exogenous forecasts}

\begin{figure}[h]
\centering
\includegraphics[width=0.75\textwidth]{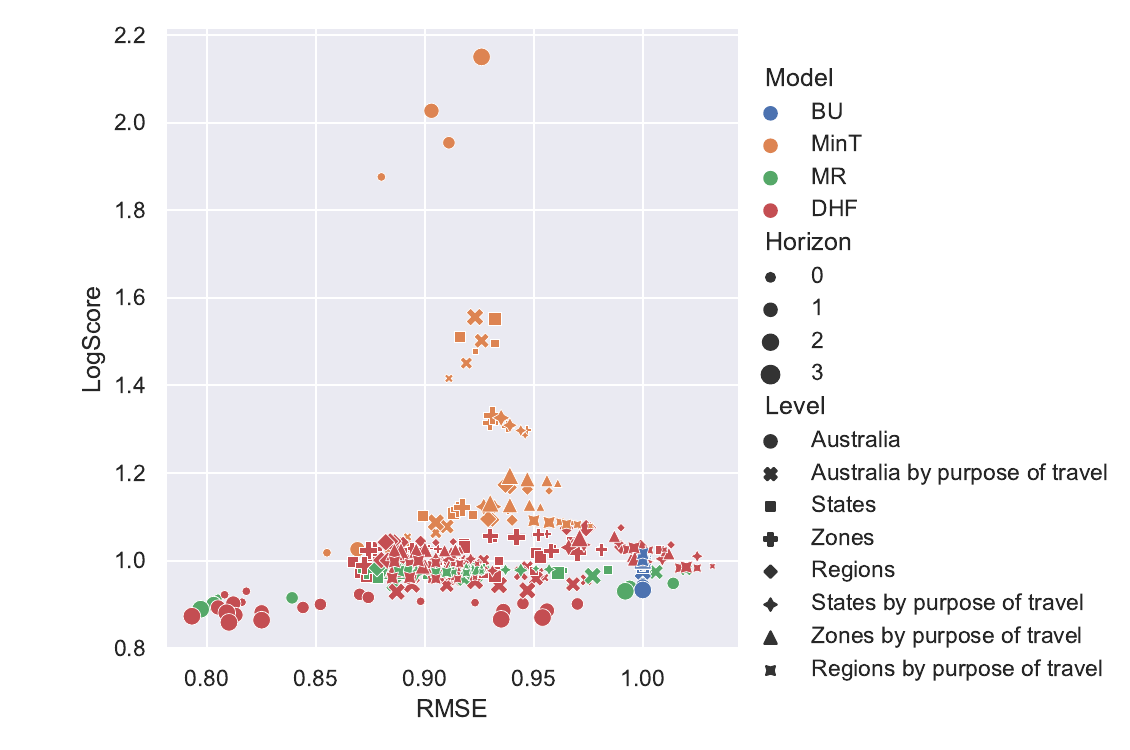}
\caption{We display Log Score (y axis) and RMSE (x axis) relative to the base case of bottom up aggregate forecasts. For full discussion of the results see main text. }
\label{fig:DHF Tourism Forecast Accucary}
\end{figure}

We now extend the MRDLM models to incorporate additional forecast information in the form of exogenous forecasts. We include these via dynamic forecast combination regressions as described above. The initial priors for the regression weights are set to $[0,(1/(2j))^2]$ where j in the number of regressors and corresponds to the number of layers in the hierarchy. For the tourism dataset, $j=8$ so that each regression weight has an initial prior of $0$ with a standard deviation of 6.25\%.  

\begin{figure}[h]
\centering
\includegraphics[width=0.75\textwidth]{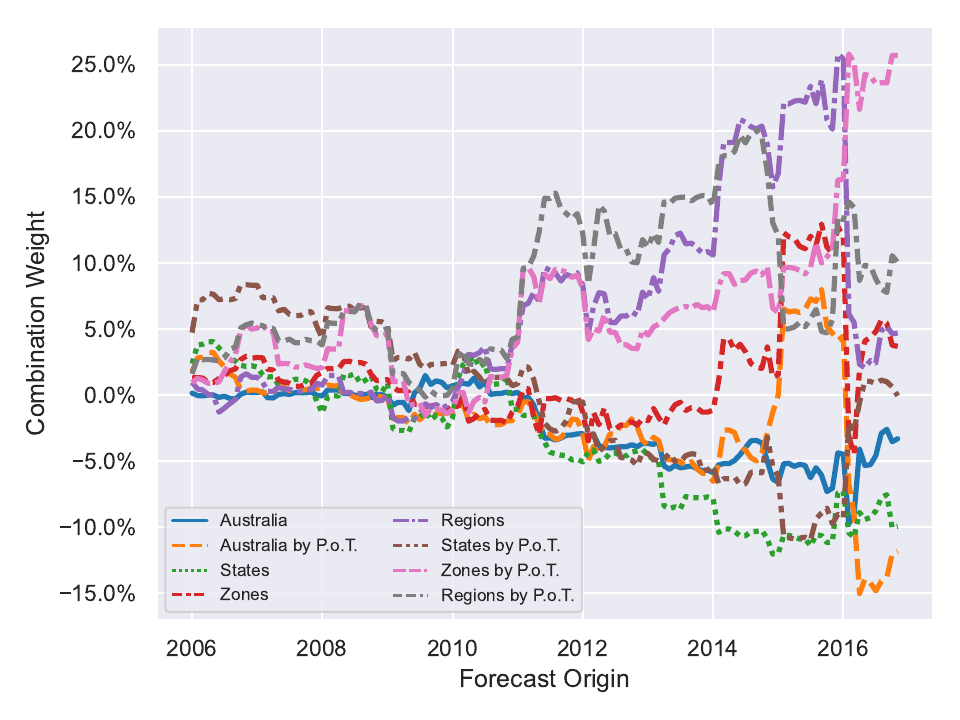}
\caption{Weight evolution over time. The chart shows the weight allocated to the forecast derived from each hierarchical layer for the North Coast, New South Wales, Holiday travel time series.}
\label{fig:Wt NSWNC Hols}
\end{figure}

We illustrate our results using a 'fast' and 'slow' dynamic regression models using regression discount rates of .97 and .99 respectively; when combined with the Fast/Medium/Slow MR-DLM models this gives rise to 6 potential model variants. The results of this exercise are set out in tabular form in Appendix B, and are summarised in figure \ref{fig:DHF Tourism Forecast Accucary}. Several features are apparent. We note in general a broadly linear relationship between RMSE and Log score for our MRDLM and DHF models. Incorporating base forecasts seems to have a beneficial effect on both point and probabilistic forecast accuracy for top level forecasts, and on RMSE more or less consistently, with the red coloured DHF forecasts dominating the bottom left hand corner of the chart, representing the largest increases in particular in distributional forecast accuracy. Where the 'Fast' MRDLM forecasts are relatively poor, including base forecasts, especially when 'Fast' rates of adaptation are allowed, improves matters significantly. We see in these cases that incorporating base forecasts mitigates against miss-specification of the MRDLM model.

It is instructive to examine the weights allocated to the forecasts derived from the various hierarchical levels. The model developed in this paper allows these to evolve dynamically over time, and potentially react to changes in circumstances more quickly than rolling or expanding window approaches. Figure \ref{fig:Wt NSWNC Hols} shows the evolution of combination weights for the largest sub series of the travel dataset, that for Holiday travel to North Coast, New South Wales. The model allocates relatively low weights to all the exogenous forecasts for the first three or so years of the sample, up until 2010, indicating that the prior MRDLM model is doing a reasonable job of forecasting. After 2010, the exogenous forecast come increasingly in to play, with geographical factors increasing in influence. This pattern broadly continues until around 2015, when the orange line, representing in this case National, holiday focused travel, increases in influence for roughly a twelve month period before falling away quite rapidly. We observe that forecasts derived from two hierarchical levels, namely the 'Total' and 'State' time series receive small negative weights on average, acting to reduce forecast variance as is relatively common in forecast combination problems. 

\begin{figure}[h]
\centering
\includegraphics[width=0.75\textwidth]{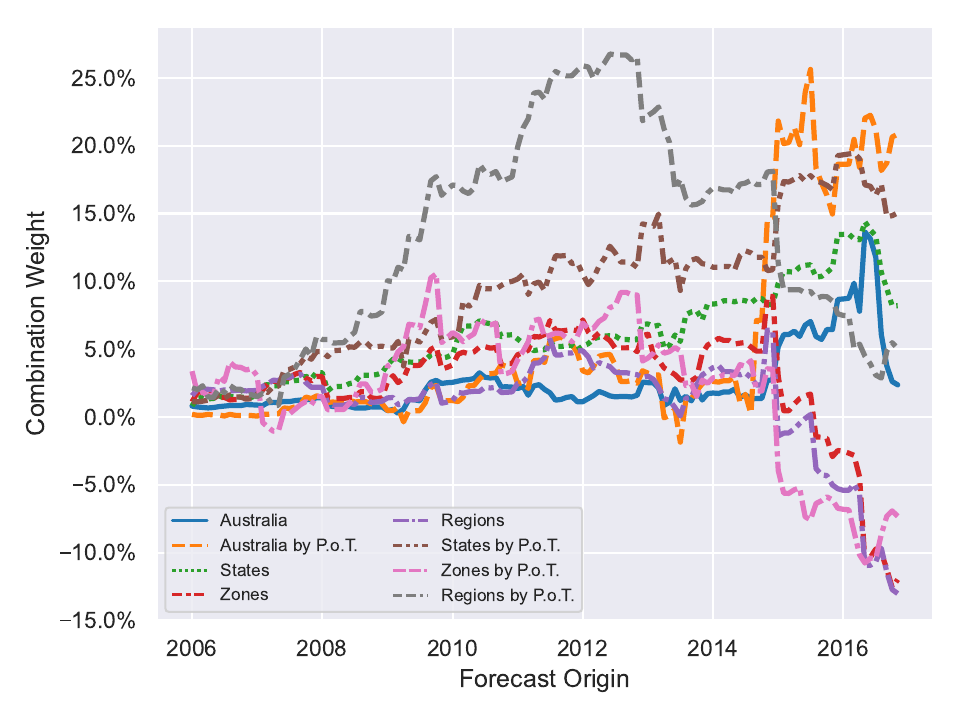}
\caption{Weight evolution over time. The chart shows the weight allocated to the forecast derived from each hierarchical layer for the Sydney Business travel time series.}
\label{fig:Wt Sydney Bus}
\end{figure}

Figure \ref{fig:Wt Sydney Bus} shows the evolution of combination weights for the largest geographical sub series of business related travel, namely that for the City of Sydney. This series demonstrates an increasing weight over time on the exogenous forecast at the base level - basically a simple forecast combination between the prior and external forecast, for the first half of the sample period. After 2012, the weight allocated to this forecast drops back somewhat. We also note a dramatic spike in the influence of forecasts for national business related travel in 2015 (orange line), weights on this factor increase rapidly at the expense of more local factors which become negative to compensate.

These charts illustrate the ability of our approach to adopt quickly to changes in dynamics. We are also able to be more transparent in terms of drawing out changes in forecast contribution than MINT based approaches, where the weights are significantly more numerous (one for each forecast in the collection) and represent rather complex linear combinations of series as described in (\cite{Hollyman2021-ek}).

\subsection{Further extensions - 2 step reconciliation and Hierarchical Combination Priors}

We now select the Medium / Fast DHF model from the above analysis, and fit 2 step and hierarchical prior versions of the same basic model. The results of this exercise are set out in the tables in Appendix C.

For the 2 step forecasts, we the States by purpose of travel level as the boundary layer between the top and bottom sub-hierarchies. We assign the exogenous forecasts derived from the Total, Total by purpose of travel, State, and State by purpose of travel time series to the top sub-hierarchy, and the remaining series to the bottom sub-hierarchies. We use the same initial priors on regression weights as above ($[0,(1/(2j))^2]$). The point forecast accuracy is very similar to that of the one step reconciliation process, showing a small deterioration for the top level, but minor improvements at the bottom level. For probabilistic forecasts, the 2 step procedure shows reasonably consistent improvement in forecast accuracy at all hierarchical levels.

As regards the hierarchical forecasts, we run these with the same prior structure on the hierarchy wide base series regressions as used above ($[0,(1/(2j))^2]$), but use somewhat tighter priors ($[0,(1/(8j))^2]$) on the extent to which series level weights deviate from the mean of all base series in the hierarchy. The empirical results are again set out in Appendix C, and show that the hierarchical models mildly under perform in terms of point forecast accuracy on this data set, although they are the best performing in terms of probabilistic accuracy across all levels of the hierarchy except for the top level.

In general terms, the substantial reduction in number of parameters, and the dynamic nature of our models leads to creditable performance on the Tourism data set and generating clear improvements, especially in terms of probabilistic forecast accuracy. We now turn to a separate forecasting experiment on the somewhat larger M5 data set.

\section{Empirical application - M5 Dataset}

The M5 data set (\cite{Makridakis2020-lk}) is roughly 75 times larger than the Tourism data set in terms of number of time series, or 660 times larger in terms of total number of time series observations. It is still somewhat 'small' relative to the problem set confronted by many multi-line retail businesses, as we describe above, but still represents a stiff computational challenge.  For relatively small hierarchies, (such as the Tourism data set) it is possible to wrap up the MRDLM model and the reconciliation regression process in one iteration. To scale the model to larger datasets, it is helpful to split the process in to two stages, running the prior model in a single process and then the dis-aggregation / combination step in a second stage as we do here. 

The M5 data comprises sales data on 3,049 products sold by Walmart in the USA, organised in to a grouped hierarchical structure with 12 levels, resulting in 30,490 base series and 42,840 series in total. The total dataset comprises 1,941 daily time series observations spanning the period from 29/1/2011 to 22/5/2016. We design the forecasting exercise to illustrate the power of our reconciliation approach, rather than forecast accuracy in the absolute sense. We therefore utilise relatively simple exogenous forecasting models, rather than trying to extract all the available information for each time series. The dataset includes information on pricing, holidays and promotional activities, which we do not attempt to incorporate in to prior or base forecasts. Doing so in a structured (Bayesian) way has the potential to substantially improve the results we present below. 

We mirror the structure used above, and select simple DLM models for the base series, incorporating level, weekly seasonality and factor regression components. For the factors we specify DLM models with level, trend and weekly seasonality components. The data also shows evidence of more complex monthly and annual seasonality which could be incorporated in to the models, but again we do not do so here.

The actual M5 competition split utilised the majority of the data for model training, and only the last 28 days for evaluation of forecast accuracy. We proceed differently, and run a quasi out of sample forecasting exercise on a weekly basis. We use the first year of daily data to produce initial prior and base forecasts, with the first set of forecasts being made as at the 28th of January 2012. The prior or baseline model is then updated sequentially in the usual way based on daily data. We utilise the fitted state space models to produce 7 step ahead prior forecasts as at the close of business on each Sunday. We produce the exogenous forecasts on the same timetable. The base forecast models are fitted using a rolling 365 day window. We re-fit the models and produce a new set of forecasts every Sunday. We note that the process of producing exogenous forecasts for the entire set of time series is extremely time consuming, even using the excellent Nixtla implementation of Rob Hyndman's Auto-ETS algorithum. \footnote{https://nixtlaverse.nixtla.io/statsforecast/src/core/models.html}

In contrast our baseline model updates rather quickly. Our experiments with the M5 data, using a desktop PC with three GPU processors will produce daily Kalman filter updates for 31,490 series in around 2 seconds. 28 step ahead forecasts for the series, including factor covariance, loading matrices and specific variances takes roughly 6 seconds. 

The reconciliation models (dis-aggregation and combination steps) are trained on the quasi out of sample exogenous and prior forecasts. We utilise a 52 week period for initial training of the reconciliation regression models, experimenting with a number of hyper-parameter specifications governing the discount rates and prior variances of the weight parameters (details available from the authors on request). We settle on the specification for the baseline model detailed in table \ref{tab:m5-hyper-parms}, and run a full quasi out of sample forecasting experiment for the remaining 176 week period, the results of which we report below.

\begin{table}[h]
\small
\centering
\caption{Hyper - Parameters M5}
\begin{tabular}{cccccccc}
\hline
\textbf{ }
\textbf{Factor}&
\textbf{Factor} &
\textbf{Base} &
\textbf{Base} &
\textbf{Base}\\
\textbf{Level/Trend}&
\textbf{Seasonality} &
\textbf{Level} &
\textbf{Seasonality} &
\textbf{Regression} &
\textbf{Variance}\\
\hline
 .99 & .995 & .995 & .997 & .997 & .9997\\
\hline
\end{tabular}
\label{tab:m5-hyper-parms}
\end{table}

For the combination regressions the sub-hierarchy wide weight parameters are set to $[0,(1/(2j))^2]$ as above for the top level hierarchy, where we do not use a hierarchical prior. The base level M5 time series are very noisy and intermittent, so we use a tighter prior structure on weights at the base series level of $[0,(1/(4j))^2]$ for the sub-hierarchy wide parameters and $[0,(1/(16j))^2]$ at series level. We use a discount rate of .99 for the daily model updates. 

We allocate the forecast information from time series comprising the M5 data set to the top or bottom hierarchies as follows:

\begin{table}[h]
\footnotesize
\centering
\caption{M5 data set time series}
\begin{tabular}{llrr}
\hline
\textbf{ }
\textbf{Level}&
\textbf{Description} &
\textbf{Number of Series} &
\textbf{Hierarchy}\\
\hline
 1 & Sales of all products, aggregated for all stores/states & 1 & Top\\
 2 & Sales of all products, aggregated for each State & 3 & Top\\
 3 & Sales of all products, aggregated for each store & 10& Top\\
 4 & Sales of all products, aggregated for each category & 3& Top\\
 5 & Sales of all products, aggregated for each department & 7& Top\\
 6 & Sales of all products, aggregated for each State and category & 9& Top\\
 7 & Sales of all products, aggregated for each State and department & 21& Top\\
 8 & Sales of all products, aggregated for each store and category & 30& Top\\
 9 & Sales of all products, aggregated for each store and department & 70& Bottom\\
 10 & Sales of product x, aggregated for all stores/states &3,075& Bottom\\
 11 & Sales of product x, aggregated for each State & 9,225& Bottom\\
 12 & Sales of product x, aggregated for each store & 30,750& Bottom\\
\hline
\end{tabular}
\label{tab:m5-series}
\end{table}

To structure our sub hierarchies, we choose as the intermediate layer time series representing 'Departmental sales by store', resulting in 70 bottom level sub-hierarchies. Due to the grouping chosen for the M5 data set, There is no natural way of imposing the regression constraints on levels 10 and 11 of the hierarchy, without using an additional set of sub-hierarchies. This could for example be achieved by adding 3075 further sub-hierarchies in a third layer, and reconciling these, passing these results to the existing sub-hierarchies as new forecasts for the base level series. We do not do that here, and instead incorporate forecast information from levels 10 and 11 into the second layer hierarchies. In practise, analysts will normally have some latitude as to which sets of time series to forecast, so we do not believe this to be a significant barrier to implementation of our model, ultimately empirical results justify our assumptions (or otherwise).

Our objective in this paper is to design a methodology which makes reconciliation algorithms available at scale, that is to real life datasets that are potentially many hundreds of times larger than the subset of the Walmart database which constitutes the M5 competition data. Existing algorithms do not scale to such datasets, as it is not feasible from a practical (let alone statistical) perspective to estimate these models due to the size of the covariance matrices involved. Even if attention is restricted to simpler MINT approaches with diagonal covariance matrices, a (non-diagonal) matrix $\mathbf{S}'\mathbf{W}_{t+h}\mathbf{S}$ needs to be inverted in order to reconcile an entire hierarchy. We note however that when the M5 competition was run, simple exponential smoothing with bottom up aggregation outperformed 92\% of the competition entrants in terms of point forecast accuracy. This model is simple to calculate and we choose to include it as our benchmark. Other alternatives might include the application of a sub-hierarchy based procedure to MINT style algorithms but we do not pursue this option here. The results of our forecasting exercise for M5 data is set out in Appendix D.

In terms of point forecast accuracy, our results broadly support the conclusions of \cite{Makridakis2022-bp} - bottom up Automatic Exponential Smoothing is a tough benchmark to beat on this dataset. These authors report that 92.5\% of the submissions in the M5 forecasting competition failed to outperform this benchmark. As described above, the rules of the M5 competition specified a different forecasting exercise to the one we conduct here, requiring a longer (28 day horizon) but a much shorter evaluation period than the 172 weeks we utilise. Our DHF models out perform AETS-BU relatively consistently across the bottom layers of the hierarchy, but struggle to do so consistently towards the top levels of the hierarchy, although the results are in general fairly close. Conversely, our baseline DHF models are a little more successful towards the top of the hierarchy, and a little less so at bottom levels. Having said this, our models are reasonably effective in improving forecasts at the important bottom or base series level, which from the vast majority of series by number, and where most decision making regarding inventory planning etc will be made, typically generating around 2\% improvement in forecast accuracy. As outlined above, we do not utilise all the M5 information in this experiment, there is also room to further optimise our reconciliation approach by accounting for information at different forecast horizons.

As outlined above, there are a number of candidate methodologies for assessing probabilistic forecast accuracy in hierarchical settings. For our purposes, methodologies based on sampling again prove to be too slow to be of practical use. We therefore, consistent with our emphasis on first and second moments in this paper, choose a scoring rule set out in \cite{Gneiting2007-ya} based on the earlier work of \cite{Dawid-dw} and \cite{Dawid1999-fq}, which focuses on data summarised by mean and variance / covariance matrix. With Gaussian data this metric collapses to the log - score.

The score is written as:

\begin{align}
    S(P, \mathbf{x}) = - \log  \det \boldsymbol{\Sigma}_P - (x -\boldsymbol{\mu}_P)' \boldsymbol{\Sigma^{-1}}_P (x - \boldsymbol{\mu}_P)
\end{align}

This is computationally relatively cheap to calculate if the estimated covariance matrix $\Sigma$ has quadratic form (via the Woodbury formula) or is diagonal.

In terms of probabilistic forecast accuracy, our reconciled DHF forecasts perform very creditably, achieving a clean sweep in terms of probabilistic forecast accuracy across the bottom 8 layers of the hierarchy, improving by circa 20\% on the benchmark bottom up diagonal approach. These improvements broadly continue towards the top of the hierarchy , where the MRDLM models also perform well. The bottom up model is inferior to both the DHF and MRDLM models for every time horizon and hierarchical level.

\section{Conclusions}

In this paper we present a revised approach to forecast reconciliation, and bring together a number of refinements which extend the benefits of this technique to large and very large hierarchical systems. We achieve significant improvements in accuracy on the benchmark Tourism data set, where the dynamic nature of our model improves on forecasting accuracy. Using the M5 data we show how our approach scales effectively and generates significant improvements in probabilistic forecast accuracy.

Room for improvement in our algorithm remains in several areas. Firstly, we need to more thoroughly model the discreet and intermittent time series which make up the bottom level series, as opposed to simply modelling their means and variances. Secondly, our experimental work suggests significant opportunities to improve our results by incorporating multi step ahead forecast errors in to our combining regressions. In addition, there remains scope to improve the efficiency of the dis-aggregation and forecasting components of our algorithm, especially when clusters of computers are available. We plan to return to these topics in future work.

\newpage
\section{Appendix A - MR-DLM Forecast Accuracy - Tourism}

\begin{table}[H]
\tiny
\resizebox{\textwidth}{!}{%
\begin{tabular}{llrrrrrrrr}
\toprule
& & \multicolumn{8}{c}{Forecast horizon (Quarters)} \\
 \cmidrule(lr){3-10}
& & \multicolumn{4}{c}{All purposes} & \multicolumn{4}{c}{By purpose of travel}\\
 \cmidrule(lr){3-6}
 \cmidrule(lr){7-10}
Level / Model &&              1 &              2 &              3 &              4 &              1 &              2 &              3 &              4 \\
\midrule
\parbox[t]{2mm}{\multirow{8}{*}{\rotatebox[origin=c]{90}{Australia}}}

&BU-Diag       &          100.0 &          100.0 &          100.0 &          100.0 &          100.0 &          100.0 &          100.0 &          100.0 \\
&BU-Shrink     &          100.0 &          100.0 &          100.0 &          100.0 &          100.0 &          100.0 &          100.0 &          100.0 \\
&MinT-OLS      &           85.4 &           89.2 &           87.3 &           88.2 &           90.3 &           91.5 &           91.6 &           89.8 \\
&MinT-WLS      &           88.0 &           91.1 &           90.3 &           92.6 &           91.1 &           91.9 &           92.6 &           92.3 \\
&MinT-Shrink   &           85.5 &           88.4 &           86.9 &           89.0 &  \textbf{89.2} &  \textbf{90.5} &  \textbf{91.0} &           90.5 \\
&MR-DLM-Fast   &           91.3 &          100.1 &           92.3 &           91.1 &          104.4 &          108.1 &          107.9 &          105.0 \\
&MR-DLM-Medium &  \textbf{80.5} &  \textbf{83.9} &  \textbf{80.3} &  \textbf{79.7} &           90.5 &           92.4 &           91.8 &  \textbf{88.6} \\
&MR-DLM-Slow   &           96.7 &          101.4 &           99.4 &           99.2 &          100.4 &          102.0 &          100.6 &           97.7 \\

\midrule
\parbox[t]{2mm}{\multirow{8}{*}{\rotatebox[origin=c]{90}{States}}}
&BU-Diag       &          100.0 &          100.0 &          100.0 &          100.0 &          100.0 &          100.0 &          100.0 &          100.0 \\
&BU-Shrink     &          100.0 &          100.0 &          100.0 &          100.0 &          100.0 &          100.0 &          100.0 &          100.0 \\
&MinT-OLS      &           92.3 &           92.6 &           90.8 &           91.3 &           95.1 &           95.1 &           94.0 &           93.8 \\
&MinT-WLS      &           92.3 &           93.2 &           91.6 &           93.2 &           94.6 &           94.4 &           93.9 &           93.5 \\
&MinT-Shrink   &           91.5 &  \textbf{92.2} &           89.9 &           91.3 &           94.0 &           93.9 &           93.2 &           92.7 \\
&MR-DLM-Fast   &          111.8 &          118.0 &          112.5 &          115.6 &          115.7 &          118.3 &          117.2 &          117.4 \\
&MR-DLM-Medium &  \textbf{91.4} &           92.7 &  \textbf{87.8} &  \textbf{87.8} &  \textbf{92.9} &  \textbf{93.7} &  \textbf{91.9} &  \textbf{89.6} \\
&MR-DLM-Slow   &           96.4 &           98.4 &           96.1 &           96.1 &           95.1 &           95.7 &           94.4 &           92.0 \\

\midrule
\parbox[t]{2mm}{\multirow{8}{*}{\rotatebox[origin=c]{90}{Zones}}}
&BU-Diag       &          100.0 &          100.0 &          100.0 &          100.0 &          100.0 &          100.0 &          100.0 &          100.0 \\
&BU-Shrink     &          100.0 &          100.0 &          100.0 &          100.0 &          100.0 &          100.0 &          100.0 &          100.0 \\
&MinT-OLS      &           95.3 &           93.6 &           92.6 &           92.2 &           97.0 &           96.5 &           95.3 &           94.3 \\
&MinT-WLS      &           94.7 &           93.8 &           93.0 &           93.1 &           96.1 &           95.6 &           94.7 &           93.9 \\
&MinT-Shrink   &           93.8 &           92.9 &           91.5 &           91.7 &           95.3 &           94.8 &           93.9 &           93.0 \\
&MR-DLM-Fast   &          114.3 &          113.9 &          111.8 &          114.6 &          116.8 &          116.1 &          114.6 &          115.1 \\
&MR-DLM-Medium &  \textbf{91.2} &  \textbf{90.2} &  \textbf{88.2} &  \textbf{87.3} &  \textbf{92.2} &  \textbf{91.9} &  \textbf{90.5} &  \textbf{88.5} \\
&MR-DLM-Slow   &           92.7 &           92.2 &           91.1 &           90.0 &           92.6 &           92.4 &           91.2 &           89.1 \\

\midrule
\parbox[t]{2mm}{\multirow{8}{*}{\rotatebox[origin=c]{90}{Regions}}}
&BU-Diag       &          100.0 &          100.0 &          100.0 &          100.0 &          100.0 &          100.0 &          100.0 &          100.0 \\
&BU-Shrink     &          100.0 &          100.0 &          100.0 &          100.0 &          100.0 &          100.0 &          100.0 &          100.0 \\
&MinT-OLS      &           96.7 &           95.0 &           94.3 &           93.4 &           98.3 &           97.8 &           97.3 &           96.8 \\
&MinT-WLS      &           95.7 &           94.7 &           93.9 &           93.7 &           97.6 &           97.0 &           96.5 &           95.7 \\
&MinT-Shrink   &           95.0 &           94.0 &           93.1 &           92.9 &           96.8 &           96.2 &           95.7 &           95.0 \\
&MR-DLM-Fast   &          118.0 &          115.3 &          114.5 &          116.4 &          119.3 &          117.4 &          116.5 &          117.2 \\
&MR-DLM-Medium &  \textbf{91.6} &  \textbf{90.1} &  \textbf{89.0} &  \textbf{87.8} &           92.6 &           91.9 &           91.0 &           89.2 \\
&MR-DLM-Slow   &  \textbf{91.6} &           90.5 &           89.9 &           88.7 &  \textbf{92.0} &  \textbf{91.3} &  \textbf{90.5} &  \textbf{88.8} \\

\bottomrule
\end{tabular}}
\caption{MR-DLM Point forecast accuracy. The left hand columns represent data grouped geographically. Right hand side columns set out data for the same geographies split by purpose of travel. The numbers are RMSE representing the forecast error for the method in question expressed as a \% of the Bottom Up aggregated base forecasts. Forecasts are made monthly, and the numbers in the table represent averages across 4 forecast quarters. Bold entries identify the best performing approaches.}
\label{MR-DLM-rmse}
\normalsize
\end{table}

\begin{table}[H]
\tiny
\resizebox{\textwidth}{!}{%
\begin{tabular}{llrrrrrrrr}
\toprule
& & \multicolumn{8}{c}{Forecast horizon (Quarters)} \\
 \cmidrule(lr){3-10}
& & \multicolumn{4}{c}{All purposes} & \multicolumn{4}{c}{By purpose of travel}\\
 \cmidrule(lr){3-6}
 \cmidrule(lr){7-10}
Level / Model &&              1 &              2 &              3 &              4 &              1 &              2 &              3 &              4 \\
\midrule
\parbox[t]{2mm}{\multirow{8}{*}{\rotatebox[origin=c]{90}{Australia}}}
&BU-Diag       &          100.0 &          100.0 &          100.0 &          100.0 &          100.0 &          100.0 &          100.0 &          100.0 \\
&BU-Shrink     &           94.9 &           94.7 &           94.1 &           93.3 &           98.0 &           98.0 &           97.8 &           97.5 \\
&MinT-OLS      &     N/M&     N/M &     N/M &     N/M &      N/M &      N/M &      N/M &     N/M \\
&MinT-WLS      &          187.6 &          195.4 &          202.7 &          215.0 &          141.6 &          145.1 &          150.2 &          155.6 \\
&MinT-Shrink   &          101.8 &          102.6 &          102.6 &          103.5 &          105.4 &          106.4 &          107.8 &          108.7 \\
&MR-DLM-Fast   &           92.8 &           92.3 &  \textbf{90.1} &  \textbf{88.5} &           99.2 &           98.5 &           97.2 &           95.6 \\
&MR-DLM-Medium &  \textbf{91.4} &  \textbf{91.5} &  \textbf{90.1} &           88.9 &  \textbf{96.4} &  \textbf{96.3} &  \textbf{95.8} &  \textbf{94.8} \\
&MR-DLM-Slow   &           94.2 &           94.8 &           93.8 &           93.0 &           97.8 &           97.9 &           97.4 &           96.5 \\

\midrule
\parbox[t]{2mm}{\multirow{8}{*}{\rotatebox[origin=c]{90}{States}}}
&BU-Diag       &          100.0 &          100.0 &          100.0 &          100.0 &          100.0 &          100.0 &          100.0 &          100.0 \\
&BU-Shrink     &           98.9 &           98.8 &           98.7 &           98.5 &           99.7 &           99.7 &           99.6 &           99.5 \\
&MinT-OLS      &      N/M&      N/M &      N/M &      N/M &      N/M &      N/M &      N/M &      N/M \\
&MinT-WLS      &          147.8 &          149.6 &          151.1 &          155.2 &          128.6 &          129.7 &          130.9 &          132.6 \\
&MinT-Shrink   &          110.3 &          110.5 &          110.2 &          110.8 &          111.5 &          112.0 &          112.3 &          112.3 \\
&MR-DLM-Fast   &          102.5 &          102.0 &          101.3 &          100.7 &          100.2 &           99.8 &           99.3 &           98.7 \\
&MR-DLM-Medium &  \textbf{97.5} &  \textbf{97.3} &  \textbf{96.7} &  \textbf{96.3} &  \textbf{98.1} &  \textbf{97.9} &  \textbf{97.6} &  \textbf{97.1} \\
&MR-DLM-Slow   &           97.8 &           97.8 &           97.5 &           97.2 &           98.2 &           98.0 &           97.8 &           97.4 \\

\midrule
\parbox[t]{2mm}{\multirow{8}{*}{\rotatebox[origin=c]{90}{Zones}}}

&BU-Diag       &          100.0 &          100.0 &          100.0 &          100.0 &          100.0 &          100.0 &          100.0 &          100.0 \\
&BU-Shrink     &           99.8 &           99.7 &           99.6 &           99.6 &          100.4 &          100.4 &          100.3 &          100.3 \\
&MinT-OLS      &     N/M &      N/M &      N/M &      N/M &       N/M &       N/M &       N/M &       N/M \\
&MinT-WLS      &          129.8 &          130.7 &          131.5 &          133.1 &          117.7 &          118.3 &          118.6 &          119.4 \\
&MinT-Shrink   &          111.6 &          111.7 &          111.7 &          112.1 &          112.3 &          112.7 &          112.7 &          113.1 \\
&MR-DLM-Fast   &          101.3 &          100.8 &          100.5 &          100.3 &          100.3 &           99.9 &           99.7 &           99.3 \\
&MR-DLM-Medium &  \textbf{98.3} &  \textbf{98.1} &  \textbf{97.9} &  \textbf{97.7} &  \textbf{98.3} &           98.2 &  \textbf{98.1} &  \textbf{97.7} \\
&MR-DLM-Slow   &  \textbf{98.3} &  \textbf{98.1} &           98.0 &           97.8 &  \textbf{98.3} &  \textbf{98.1} &  \textbf{98.1} &  \textbf{97.7} \\

\midrule
\parbox[t]{2mm}{\multirow{8}{*}{\rotatebox[origin=c]{90}{Regions}}}
&BU-Diag       &          100.0 &          100.0 &          100.0 &          100.0 &          100.0 &          100.0 &          100.0 &          100.0 \\
&BU-Shrink     &          100.1 &          100.1 &          100.1 &          100.1 &          101.5 &          101.6 &          101.6 &          101.6 \\
&MinT-OLS      &       N/M &       N/M &       N/M.3 &      N/M &       N/M &       N/M &       N/M &       N/M \\
&MinT-WLS      &          115.9 &          116.3 &          116.7 &          117.3 &          107.9 &          108.2 &          108.3 &          108.7 \\
&MinT-Shrink   &          109.2 &          109.2 &          109.3 &          109.5 &          108.6 &          108.8 &          108.9 &          109.1 \\
&MR-DLM-Fast   &          101.6 &          101.2 &          100.9 &          100.9 &           99.4 &           98.9 &           98.8 &           99.2 \\
&MR-DLM-Medium &           98.7 &  \textbf{98.5} &  \textbf{98.3} &           98.3 &  \textbf{97.5} &  \textbf{97.2} &  \textbf{97.3} &  \textbf{97.6} \\
&MR-DLM-Slow   &  \textbf{98.6} &  \textbf{98.5} &  \textbf{98.3} &  \textbf{98.2} &  \textbf{97.5} &  \textbf{97.2} &  \textbf{97.3} &  \textbf{97.6} \\

\bottomrule
\end{tabular}}
\caption{MR-DLM Probabilistic forecast accuracy. Layout as per table \ref{MR-DLM-rmse}. The numbers are log scores representing the forecast error for the method in question expressed as a \% of the Bottom Up aggregated base forecasts. Bold entries identify the best performing approaches. Results for MinT - OLS have log scores which are not meaningful (N/M)}
\label{MR-DLM-logscore}
\normalsize
\end{table}

\newpage
\section{Appendix B - DHF Forecast Accuracy - Tourism}

\begin{table}[H]
\footnotesize
\resizebox{\textwidth}{!}{%
\begin{tabular}{llrrrrrrrr}
\toprule
& & \multicolumn{8}{c}{Forecast horizon (Quarters)} \\
 \cmidrule(lr){3-10}
& & \multicolumn{4}{c}{All purposes} & \multicolumn{4}{c}{By purpose of travel}\\
 \cmidrule(lr){3-6}
 \cmidrule(lr){7-10}
Level / Model &&              1 &              2 &              3 &              4 &              1 &              2 &              3 &              4 \\
\midrule
\parbox[t]{2mm}{\multirow{13}{*}{\rotatebox[origin=c]{90}{Australia}}}
&BU-Diag         &          100.0 &          100.0 &          100.0 &          100.0 &          100.0 &          100.0 &          100.0 &          100.0 \\
&BU-Shrink       &          100.0 &          100.0 &          100.0 &          100.0 &          100.0 &          100.0 &          100.0 &          100.0 \\
&MinT-Shrink     &           85.5 &           88.4 &           86.9 &           89.0 &           89.2 &  \textbf{90.5} &  \textbf{91.0} &           90.5 \\
&MinT-WLS        &           88.0 &           91.1 &           90.3 &           92.6 &           91.1 &           91.9 &           92.6 &           92.3 \\
&MR-DLM-Fast     &           88.2 &           98.5 &           90.1 &           89.0 &          100.1 &          105.1 &          105.3 &          102.5 \\
&MR-DLM-Medium   &  \textbf{80.5} &  \textbf{83.9} &  \textbf{80.3} &           79.7 &           90.5 &           92.4 &           91.8 &  \textbf{88.6} \\
&MR-DLM-Slow     &           96.7 &          101.4 &           99.4 &           99.2 &          100.4 &          102.0 &          100.6 &           97.7 \\
&DHF-Fast/Fast   &           81.8 &           87.0 &           81.2 &           80.9 &           89.8 &           92.7 &           92.3 &           91.0 \\
&DHF-Fast/Slow   &           80.8 &           87.4 &           80.5 &  \textbf{79.3} &           91.3 &           94.7 &           94.4 &           92.3 \\
&DHF-Medium/Fast &           81.6 &           85.2 &           82.5 &           82.5 &  \textbf{88.7} &           91.2 &           91.1 &           89.4 \\
&DHF-Medium/Slow &           81.0 &           84.4 &           81.3 &           81.0 &           89.2 &           91.4 &  \textbf{91.0} &           88.7 \\
&DHF-Slow/Fast   &           89.8 &           94.5 &           93.6 &           93.5 &           93.0 &           95.2 &           95.1 &           93.4 \\
&DHF-Slow/Slow   &           92.3 &           97.0 &           95.6 &           95.4 &           95.5 &           97.4 &           96.8 &           94.7 \\

\midrule
\parbox[t]{2mm}{\multirow{13}{*}{\rotatebox[origin=c]{90}{States}}}

\bottomrule
\end{tabular}}
\caption{DHF point forecast accuracy - upper levels. Layout as per table \ref{MR-DLM-rmse}. The numbers are RMSE for the method in question expressed as a \% of the Bottom up aggregated base forecasts. Bold entries identify the best performing approaches}
\label{DHF-rmse-top}
\end{table}

\begin{table}[H]
\footnotesize
\resizebox{\textwidth}{!}{%
\begin{tabular}{llrrrrrrrr}
\toprule
& & \multicolumn{8}{c}{Forecast horizon (Quarters)} \\
 \cmidrule(lr){3-10}
& & \multicolumn{4}{c}{All purposes} & \multicolumn{4}{c}{By purpose of travel}\\
 \cmidrule(lr){3-6}
 \cmidrule(lr){7-10}
Level / Model &&              1 &              2 &              3 &              4 &              1 &              2 &              3 &              4 \\
\midrule
\parbox[t]{2mm}{\multirow{13}{*}{\rotatebox[origin=c]{90}{Zones}}}

&BU-Diag         &          100.0 &          100.0 &          100.0 &          100.0 &          100.0 &          100.0 &          100.0 &          100.0 \\
&BU-Shrink       &          100.0 &          100.0 &          100.0 &          100.0 &          100.0 &          100.0 &          100.0 &          100.0 \\
&MinT-Shrink     &           93.8 &           92.9 &           91.5 &           91.7 &           95.3 &           94.8 &           93.9 &           93.0 \\
&MinT-WLS        &           94.7 &           93.8 &           93.0 &           93.1 &           96.1 &           95.6 &           94.7 &           93.9 \\
&MR-DLM-Fast     &          107.3 &          108.3 &          106.1 &          108.3 &          108.5 &          109.1 &          107.7 &          107.8 \\
&MR-DLM-Medium   &           91.2 &           90.2 &           88.2 &           87.3 &           92.2 &           91.9 &           90.5 &  \textbf{88.5} \\
&MR-DLM-Slow     &           92.7 &           92.2 &           91.1 &           90.0 &           92.6 &           92.4 &           91.2 &           89.1 \\
&DHF-Fast/Fast   &           95.6 &           95.2 &           93.0 &           94.2 &           98.7 &           98.7 &           97.1 &           97.1 \\
&DHF-Fast/Slow   &           98.1 &           98.1 &           95.8 &           97.0 &          101.0 &          101.2 &           99.6 &           99.7 \\
&DHF-Medium/Fast &  \textbf{90.2} &  \textbf{89.4} &  \textbf{87.7} &           87.4 &           91.6 &           91.4 &           90.0 &           88.9 \\
&DHF-Medium/Slow &           90.5 &           89.5 &  \textbf{87.7} &  \textbf{87.2} &           91.6 &           91.4 &           90.0 &           88.6 \\
&DHF-Slow/Fast   &           90.6 &           89.9 &           88.8 &           88.8 &  \textbf{91.2} &  \textbf{90.9} &  \textbf{89.6} &           88.6 \\
&DHF-Slow/Slow   &           91.3 &           90.5 &           89.6 &           89.2 &           91.5 &           91.1 &           89.9 &           88.6 \\

\midrule
\parbox[t]{2mm}{\multirow{13}{*}{\rotatebox[origin=c]{90}{Regions}}}

&BU-Diag         &          100.0 &          100.0 &          100.0 &          100.0 &          100.0 &          100.0 &          100.0 &          100.0 \\
&BU-Shrink       &          100.0 &          100.0 &          100.0 &          100.0 &          100.0 &          100.0 &          100.0 &          100.0 \\
&MinT-Shrink     &           95.0 &           94.0 &           93.1 &           92.9 &           96.8 &           96.2 &           95.7 &           95.0 \\
&MinT-WLS        &           95.7 &           94.7 &           93.9 &           93.7 &           97.6 &           97.0 &           96.5 &           95.7 \\
&MR-DLM-Fast     &          109.0 &          107.8 &          107.2 &          108.6 &          109.7 &          109.1 &          108.5 &          108.9 \\
&MR-DLM-Medium   &           91.6 &           90.1 &           89.0 &  \textbf{87.8} &           92.6 &           91.9 &           91.0 &           89.2 \\
&MR-DLM-Slow     &           91.6 &           90.5 &           89.9 &           88.7 &           92.0 &           91.3 &           90.5 &           88.8 \\
&DHF-Fast/Fast   &           99.0 &           97.6 &           96.5 &           97.4 &          101.0 &          100.3 &           99.3 &           99.6 \\
&DHF-Fast/Slow   &          101.3 &           99.9 &           99.0 &           99.9 &          103.2 &          102.5 &          101.7 &          102.0 \\
&DHF-Medium/Fast &           91.3 &           89.8 &           88.6 &           88.4 &           92.4 &           91.6 &           90.5 &           89.7 \\
&DHF-Medium/Slow &           91.3 &           89.7 &           88.6 &           88.0 &           92.3 &           91.5 &           90.5 &           89.3 \\
&DHF-Slow/Fast   &  \textbf{90.5} &  \textbf{89.0} &  \textbf{88.1} &           88.2 &  \textbf{91.3} &  \textbf{90.3} &  \textbf{89.3} &           88.6 \\
&DHF-Slow/Slow   &           90.7 &           89.2 &           88.6 &           88.3 &  \textbf{91.3} &  \textbf{90.3} &           89.4 &  \textbf{88.5} \\

\bottomrule
\end{tabular}}
\caption{DHF point forecast accuracy - lower levels. Layout as per table \ref{MR-DLM-rmse}. The numbers are RMSE for the method in question expressed as a \% of the Bottom up aggregated base forecasts. Bold entries identify the best performing approaches}
\label{DHF-rmse-bottom}
\end{table}

\begin{table}[H]
\footnotesize
\resizebox{\textwidth}{!}{%
\begin{tabular}{llrrrrrrrr}
\toprule
& & \multicolumn{8}{c}{Forecast horizon (Quarters)} \\
 \cmidrule(lr){3-10}
& & \multicolumn{4}{c}{All purposes} & \multicolumn{4}{c}{By purpose of travel}\\
 \cmidrule(lr){3-6}
 \cmidrule(lr){7-10}
Level / Model &&              1 &              2 &              3 &              4 &              1 &              2 &              3 &              4 \\
\midrule
\parbox[t]{2mm}{\multirow{13}{*}{\rotatebox[origin=c]{90}{Australia}}}
&BU-Diag         &          100.0 &          100.0 &          100.0 &          100.0 &          100.0 &          100.0 &          100.0 &          100.0 \\
&BU-Shrink       &           94.9 &           94.7 &           94.1 &           93.3 &           98.0 &           98.0 &           97.8 &           97.5 \\
&MinT-Shrink     &          101.8 &          102.6 &          102.6 &          103.5 &          105.4 &          106.4 &          107.8 &          108.7 \\
&MinT-WLS        &          187.6 &          195.4 &          202.7 &          215.0 &          141.6 &          145.1 &          150.2 &          155.6 \\
&MR-DLM-Medium   &           91.4 &           91.5 &           90.1 &           88.9 &           96.4 &           96.3 &           95.8 &           94.8 \\
&MR-DLM-Medium   &           91.4 &           91.5 &           90.1 &           88.9 &           96.4 &           96.3 &           95.8 &           94.8 \\
&MR-DLM-Slow     &           94.2 &           94.8 &           93.8 &           93.0 &           97.8 &           97.9 &           97.4 &           96.5 \\
&DHF-Fast/Fast   &           93.0 &           92.3 &           90.2 &           88.1 &          100.9 &          100.1 &           98.8 &           97.0 \\
&DHF-Fast/Slow   &           92.2 &           91.6 &           89.3 &           87.3 &           99.4 &           98.6 &           97.1 &           95.3 \\
&DHF-Medium/Fast &           90.5 &           90.0 &           88.2 &           86.4 &           97.7 &           97.1 &           96.0 &           94.5 \\
&DHF-Medium/Slow &  \textbf{89.6} &  \textbf{89.3} &  \textbf{87.6} &  \textbf{85.9} &  \textbf{95.9} &  \textbf{95.4} &  \textbf{94.4} &  \textbf{92.9} \\
&DHF-Slow/Fast   &           90.7 &           90.2 &           88.5 &           86.6 &           97.6 &           97.1 &           95.9 &           94.4 \\
&DHF-Slow/Slow   &           90.4 &           90.1 &           88.6 &           87.0 &           96.1 &           95.6 &           94.6 &           93.2 \\

\midrule
\parbox[t]{2mm}{\multirow{13}{*}{\rotatebox[origin=c]{90}{States}}}

\bottomrule
\end{tabular}}
\caption{DHF probablistic forecast accuracy - upper levels. Layout as per table \ref{MR-DLM-rmse}. The numbers are RMSE for the method in question expressed as a \% of the Bottom up aggregated base forecasts. Bold entries identify the best performing approaches}
\label{DHF-logsocre-top}
\end{table}

\begin{table}[H]
\footnotesize
\resizebox{\textwidth}{!}{%
\begin{tabular}{llrrrrrrrr}
\toprule
& & \multicolumn{8}{c}{Forecast horizon (Quarters)} \\
 \cmidrule(lr){3-10}
& & \multicolumn{4}{c}{All purposes} & \multicolumn{4}{c}{By purpose of travel}\\
 \cmidrule(lr){3-6}
 \cmidrule(lr){7-10}
Level / Model &&              1 &              2 &              3 &              4 &              1 &              2 &              3 &              4 \\
\midrule
\parbox[t]{2mm}{\multirow{13}{*}{\rotatebox[origin=c]{90}{Zones}}}

&BU-Diag         &          100.0 &          100.0 &          100.0 &          100.0 &          100.0 &          100.0 &          100.0 &          100.0 \\
&BU-Shrink       &           99.8 &           99.7 &           99.6 &           99.6 &          100.4 &          100.4 &          100.3 &          100.3 \\
&MinT-Shrink     &          111.6 &          111.7 &          111.7 &          112.1 &          112.3 &          112.7 &          112.7 &          113.1 \\
&MinT-WLS        &          129.8 &          130.7 &          131.5 &          133.1 &          117.7 &          118.3 &          118.6 &          119.4 \\
&MR-DLM-Medium   &  \textbf{98.3} &  \textbf{98.1} &  \textbf{97.9} &  \textbf{97.7} &  \textbf{98.3} &           98.2 &  \textbf{98.1} &  \textbf{97.7} \\
&MR-DLM-Medium   &  \textbf{98.3} &  \textbf{98.1} &  \textbf{97.9} &  \textbf{97.7} &  \textbf{98.3} &           98.2 &  \textbf{98.1} &  \textbf{97.7} \\
&MR-DLM-Slow     &  \textbf{98.3} &  \textbf{98.1} &           98.0 &           97.8 &  \textbf{98.3} &  \textbf{98.1} &  \textbf{98.1} &  \textbf{97.7} \\
&DHF-Fast/Fast   &          106.1 &          105.9 &          105.7 &          105.4 &          105.7 &          105.6 &          105.6 &          105.2 \\
&DHF-Fast/Slow   &          102.9 &          102.5 &          102.2 &          101.8 &          101.9 &          101.6 &          101.4 &          100.8 \\
&DHF-Medium/Fast &          102.8 &          102.7 &          102.7 &          102.4 &          102.5 &          102.5 &          102.5 &          102.1 \\
&DHF-Medium/Slow &           99.7 &           99.4 &           99.2 &           98.9 &           99.1 &           98.9 &           98.8 &           98.3 \\
&DHF-Slow/Fast   &          102.7 &          102.6 &          102.5 &          102.3 &          102.3 &          102.3 &          102.3 &          102.0 \\
&DHF-Slow/Slow   &           99.6 &           99.3 &           99.1 &           98.8 &           99.0 &           98.8 &           98.7 &           98.2 \\

\midrule
\parbox[t]{2mm}{\multirow{13}{*}{\rotatebox[origin=c]{90}{Regions}}}
&BU-Diag         &          100.0 &          100.0 &          100.0 &          100.0 &          100.0 &          100.0 &          100.0 &          100.0 \\
&BU-Shrink       &          100.1 &          100.1 &          100.1 &          100.1 &          101.5 &          101.6 &          101.6 &          101.6 \\
&MinT-Shrink     &          109.2 &          109.2 &          109.3 &          109.5 &          108.6 &          108.8 &          108.9 &          109.1 \\
&MinT-WLS        &          115.9 &          116.3 &          116.7 &          117.3 &          107.9 &          108.2 &          108.3 &          108.7 \\
&MR-DLM-Medium   &           98.7 &  \textbf{98.5} &  \textbf{98.3} &           98.3 &           97.5 &           97.2 &           97.3 &           97.6 \\
&MR-DLM-Medium   &           98.7 &  \textbf{98.5} &  \textbf{98.3} &           98.3 &           97.5 &           97.2 &           97.3 &           97.6 \\
&MR-DLM-Slow     &  \textbf{98.6} &  \textbf{98.5} &  \textbf{98.3} &  \textbf{98.2} &           97.5 &           97.2 &           97.3 &           97.6 \\
&DHF-Fast/Fast   &          107.5 &          107.5 &          107.4 &          107.4 &          102.5 &          102.4 &          102.5 &          102.9 \\
&DHF-Fast/Slow   &          103.6 &          103.3 &          103.1 &          102.9 &           98.7 &           98.3 &           98.3 &           98.5 \\
&DHF-Medium/Fast &          104.3 &          104.4 &          104.4 &          104.4 &           99.3 &           99.3 &           99.5 &           99.9 \\
&DHF-Medium/Slow &          100.5 &          100.4 &          100.2 &          100.1 &           96.1 &           95.8 &           95.8 &           96.1 \\
&DHF-Slow/Fast   &          104.1 &          104.2 &          104.2 &          104.2 &           99.2 &           99.1 &           99.3 &           99.7 \\
&DHF-Slow/Slow   &          100.4 &          100.3 &          100.1 &          100.0 &  \textbf{96.0} &  \textbf{95.7} &  \textbf{95.7} &  \textbf{96.0} \\

\bottomrule
\end{tabular}}
\caption{DHF probablistic forecast accuracy - lower levels. Layout as per table \ref{MR-DLM-rmse}. The numbers are log scores for the method in question expressed as a \% of the Bottom up aggregated base forecasts. Bold entries identify the best performing approaches}
\label{DHF-logscore-bottom}
\end{table}

\newpage
\section{Appendix C - DHF Variants Forecast Accuracy - Tourism}

\begin{table}[H]
\footnotesize
\resizebox{\textwidth}{!}{%
\begin{tabular}{llrrrrrrrr}
\toprule
& & \multicolumn{8}{c}{Forecast horizon (Quarters)} \\
 \cmidrule(lr){3-10}
& & \multicolumn{4}{c}{All purposes} & \multicolumn{4}{c}{By purpose of travel}\\
 \cmidrule(lr){3-6}
 \cmidrule(lr){7-10}
Level / Model &&              1 &              2 &              3 &              4 &              1 &              2 &              3 &              4 \\
\midrule
\parbox[t]{2mm}{\multirow{6}{*}{\rotatebox[origin=c]{90}{Australia}}}
&BU-Diag                              &          100.0 &          100.0 &          100.0 &          100.0 &          100.0 &          100.0 &          100.0 &          100.0 \\
&BU-Shrink                            &          100.0 &          100.0 &          100.0 &          100.0 &          100.0 &          100.0 &          100.0 &          100.0 \\
&MinT-WLS                             &           88.0 &           91.1 &           90.3 &           92.6 &           91.1 &           91.9 &           92.6 &           92.3 \\
&MinT-Shrink                          &           85.5 &           88.4 &           86.9 &           89.0 &           89.2 &  \textbf{90.5} &           91.0 &           90.5 \\
&DHF-Medium/Fast                      &  \textbf{81.6} &  \textbf{85.2} &  \textbf{82.5} &  \textbf{82.5} &           88.7 &           91.2 &           91.1 &  \textbf{89.4} \\
&DHF-Medium/Fast - 2 Step             &           82.0 &           85.6 &           83.1 &           83.2 &           89.0 &           91.5 &           91.4 &           89.6 \\
&DHF-Medium/Fast - Hierarchical Prior &           83.1 &           85.9 &           84.5 &           84.4 &  \textbf{88.4} &           90.8 &  \textbf{90.9} &           90.4 \\

\midrule
\parbox[t]{2mm}{\multirow{6}{*}{\rotatebox[origin=c]{90}{States}}}
&BU-Diag                              &          100.0 &          100.0 &          100.0 &          100.0 &          100.0 &          100.0 &          100.0 &          100.0 \\
&BU-Shrink                            &          100.0 &          100.0 &          100.0 &          100.0 &          100.0 &          100.0 &          100.0 &          100.0 \\
&MinT-WLS                             &           92.3 &           93.2 &           91.6 &           93.2 &           94.6 &           94.4 &           93.9 &           93.5 \\
&MinT-Shrink                          &           91.5 &           92.2 &           89.9 &           91.3 &           94.0 &           93.9 &           93.2 &           92.7 \\
&DHF-Medium/Fast                      &           89.6 &           90.8 &  \textbf{86.7} &  \textbf{87.3} &  \textbf{91.3} &  \textbf{92.1} &  \textbf{90.5} &  \textbf{89.2} \\
&DHF-Medium/Fast - 2 Step             &           89.7 &           91.0 &           87.1 &           87.7 &           91.4 &           92.2 &           90.7 &           89.3 \\
&DHF-Medium/Fast - Hierarchical Prior &  \textbf{89.2} &  \textbf{90.3} &           88.1 &           88.7 &           91.4 &           92.5 &           91.6 &           91.1 \\

\midrule
\parbox[t]{2mm}{\multirow{6}{*}{\rotatebox[origin=c]{90}{Zones}}}
&BU-Diag                              &          100.0 &          100.0 &          100.0 &          100.0 &          100.0 &          100.0 &          100.0 &          100.0 \\
&BU-Shrink                            &          100.0 &          100.0 &          100.0 &          100.0 &          100.0 &          100.0 &          100.0 &          100.0 \\
&MinT-WLS                             &           94.7 &           93.8 &           93.0 &           93.1 &           96.1 &           95.6 &           94.7 &           93.9 \\
&MinT-Shrink                          &           93.8 &           92.9 &           91.5 &           91.7 &           95.3 &           94.8 &           93.9 &           93.0 \\
&DHF-Medium/Fast                      &  \textbf{90.2} &  \textbf{89.4} &  \textbf{87.7} &           87.4 &  \textbf{91.6} &  \textbf{91.4} &  \textbf{90.0} &  \textbf{88.9} \\
&DHF-Medium/Fast - 2 Step             &           90.3 &  \textbf{89.4} &  \textbf{87.7} &  \textbf{87.3} &  \textbf{91.6} &  \textbf{91.4} &           90.1 &  \textbf{88.9} \\
&DHF-Medium/Fast - Hierarchical Prior &           90.6 &           90.2 &           88.9 &           88.7 &           91.9 &           91.9 &           90.8 &           90.2 \\

\midrule
\parbox[t]{2mm}{\multirow{6}{*}{\rotatebox[origin=c]{90}{Regions}}}
&BU-Diag                              &          100.0 &          100.0 &          100.0 &          100.0 &          100.0 &          100.0 &          100.0 &          100.0 \\
&BU-Shrink                            &          100.0 &          100.0 &          100.0 &          100.0 &          100.0 &          100.0 &          100.0 &          100.0 \\
&MinT-WLS                             &           95.7 &           94.7 &           93.9 &           93.7 &           97.6 &           97.0 &           96.5 &           95.7 \\
&MinT-Shrink                          &           95.0 &           94.0 &           93.1 &           92.9 &           96.8 &           96.2 &           95.7 &           95.0 \\
&DHF-Medium/Fast                      &  \textbf{91.3} &           89.8 &  \textbf{88.6} &           88.4 &  \textbf{92.4} &  \textbf{91.6} &  \textbf{90.5} &           89.7 \\
&DHF-Medium/Fast - 2 Step             &           91.4 &  \textbf{89.7} &           88.7 &  \textbf{88.3} &           92.5 &  \textbf{91.6} &           90.6 &  \textbf{89.6} \\
&DHF-Medium/Fast - Hierarchical Prior &           91.4 &           90.6 &           89.5 &           89.5 &           92.5 &           92.1 &           91.1 &           90.6 \\

\bottomrule
\end{tabular}}
\caption{DHF 1 step v. 2 step, and Hierarchical Prior point forecast accuracy. Layout as per table \ref{MR-DLM-rmse}. The numbers are RMSE for the method in question expressed as a \% of the Bottom up aggregated base forecasts. Bold entries identify the best performing approaches}
\label{DHF-rmse-1v2step_point}
\end{table}

\begin{table}[H]
\footnotesize
\resizebox{\textwidth}{!}{%
\begin{tabular}{llrrrrrrrr}
\toprule
& & \multicolumn{8}{c}{Forecast horizon (Quarters)} \\
 \cmidrule(lr){3-10}
& & \multicolumn{4}{c}{All purposes} & \multicolumn{4}{c}{By purpose of travel}\\
 \cmidrule(lr){3-6}
 \cmidrule(lr){7-10}
Level / Model &&              1 &              2 &              3 &              4 &              1 &              2 &              3 &              4 \\
\midrule
\parbox[t]{2mm}{\multirow{6}{*}{\rotatebox[origin=c]{90}{Australia}}}

&BU-Diag                              &          100.0 &          100.0 &          100.0 &          100.0 &          100.0 &          100.0 &          100.0 &          100.0 \\
&BU-Shrink                            &           94.9 &           94.7 &           94.1 &           93.3 &           98.0 &           98.0 &           97.8 &           97.5 \\
&MinT-WLS                             &          187.6 &          195.4 &          202.7 &          215.0 &          141.6 &          145.1 &          150.2 &          155.6 \\
&MinT-Shrink                          &          101.8 &          102.6 &          102.6 &          103.5 &          105.4 &          106.4 &          107.8 &          108.7 \\
&DHF-Medium/Fast                      &           90.5 &           90.0 &           88.2 &           86.4 &           97.7 &           97.1 &           96.0 &           94.5 \\
&DHF-Medium/Fast - 2 Step             &  \textbf{90.2} &  \textbf{89.8} &  \textbf{88.0} &  \textbf{86.2} &           97.0 &           96.4 &           95.3 &           93.8 \\
&DHF-Medium/Fast - Hierarchical Prior &           90.6 &           90.3 &           88.7 &           87.2 &  \textbf{96.1} &  \textbf{95.6} &  \textbf{94.6} &  \textbf{93.3} \\

\midrule
\parbox[t]{2mm}{\multirow{6}{*}{\rotatebox[origin=c]{90}{States}}}

&BU-Diag                              &          100.0 &          100.0 &          100.0 &          100.0 &          100.0 &          100.0 &          100.0 &          100.0 \\
&BU-Shrink                            &           98.9 &           98.8 &           98.7 &           98.5 &           99.7 &           99.7 &           99.6 &           99.5 \\
&MinT-WLS                             &          147.8 &          149.6 &          151.1 &          155.2 &          128.6 &          129.7 &          130.9 &          132.6 \\
&MinT-Shrink                          &          110.3 &          110.5 &          110.2 &          110.8 &          111.5 &          112.0 &          112.3 &          112.3 \\
&DHF-Medium/Fast                      &          100.3 &          100.1 &           99.7 &           99.1 &          100.8 &          100.6 &          100.3 &           99.8 \\
&DHF-Medium/Fast - 2 Step             &           99.4 &           99.2 &           98.7 &           98.1 &           99.6 &           99.4 &           99.1 &           98.6 \\
&DHF-Medium/Fast - Hierarchical Prior &  \textbf{97.8} &  \textbf{97.5} &  \textbf{97.0} &  \textbf{96.4} &  \textbf{97.7} &  \textbf{97.4} &  \textbf{97.0} &  \textbf{96.4} \\
\midrule
\parbox[t]{2mm}{\multirow{6}{*}{\rotatebox[origin=c]{90}{Zones}}}

&BU-Diag                              &          100.0 &          100.0 &          100.0 &          100.0 &          100.0 &          100.0 &          100.0 &          100.0 \\
&BU-Shrink                            &           99.8 &           99.7 &           99.6 &           99.6 &          100.4 &          100.4 &          100.3 &          100.3 \\
&MinT-WLS                             &          129.8 &          130.7 &          131.5 &          133.1 &          117.7 &          118.3 &          118.6 &          119.4 \\
&MinT-Shrink                          &          111.6 &          111.7 &          111.7 &          112.1 &          112.3 &          112.7 &          112.7 &          113.1 \\
&DHF-Medium/Fast                      &          102.8 &          102.7 &          102.7 &          102.4 &          102.5 &          102.5 &          102.5 &          102.1 \\
&DHF-Medium/Fast - 2 Step             &          101.4 &          101.3 &          101.2 &          100.9 &          100.8 &          100.8 &          100.8 &          100.4 \\
&DHF-Medium/Fast - Hierarchical Prior &  \textbf{99.2} &  \textbf{99.0} &  \textbf{98.7} &  \textbf{98.4} &  \textbf{98.5} &  \textbf{98.3} &  \textbf{98.2} &  \textbf{97.7} \\

\midrule
\parbox[t]{2mm}{\multirow{6}{*}{\rotatebox[origin=c]{90}{Regions}}}

&BU-Diag                              &          100.0 &          100.0 &          100.0 &          100.0 &          100.0 &          100.0 &          100.0 &          100.0 \\
&BU-Shrink                            &          100.1 &          100.1 &          100.1 &          100.1 &          101.5 &          101.6 &          101.6 &          101.6 \\
&MinT-WLS                             &          115.9 &          116.3 &          116.7 &          117.3 &          107.9 &          108.2 &          108.3 &          108.7 \\
&MinT-Shrink                          &          109.2 &          109.2 &          109.3 &          109.5 &          108.6 &          108.8 &          108.9 &          109.1 \\
&DHF-Medium/Fast                      &          104.3 &          104.4 &          104.4 &          104.4 &           99.3 &           99.3 &           99.5 &           99.9 \\
&DHF-Medium/Fast - 2 Step             &          102.5 &          102.5 &          102.5 &          102.5 &           97.9 &           97.7 &           97.9 &           98.3 \\
&DHF-Medium/Fast - Hierarchical Prior &  \textbf{99.9} &  \textbf{99.7} &  \textbf{99.5} &  \textbf{99.4} &  \textbf{96.4} &  \textbf{96.1} &  \textbf{96.2} &  \textbf{96.5} \\
\bottomrule
\end{tabular}}
\caption{DHF 1 step v. 2 step and Hierarchical Prior probabilistic forecast accuracy. Layout as per table \ref{MR-DLM-rmse}. The numbers are log scores for the method in question expressed as a \% of the Bottom up aggregated base forecasts. Bold entries identify the best performing approaches}
\label{DHF-rmse-1v2step_prob}
\end{table}

\newpage
\section{Appendix D - Forecast Accuracy - M5}

\begin{table}[H]
\footnotesize
\resizebox{\textwidth}{!}{%
\begin{tabular}{llrrrrrrrr}
\toprule
& & \multicolumn{7}{c}{Forecast horizon (Days)} \\
Level / Model &&              1 &              2 &              3 &              4 &              5 &              6 &              7 \\
\midrule
\parbox[t]{2mm}{\multirow{4}{*}{\rotatebox[origin=c]{90}{Total}}}
& AETS & \textbf{0.8438} & \textbf{0.9825} & 1.0042 & 0.9995 & 1.0084 & 1.0814 & 0.9916 \\
 & AETS-BU & 1.0000 & 1.0000 & 1.0000 & 1.0000 & 1.0000 & \textbf{1.0000} & 1.0000 \\
 & DHF & 1.2588 & 1.2289 & \textbf{0.9829} & \textbf{0.9734} & \textbf{0.9631} & 1.0249 & 1.0059 \\
 & MRDLM & 0.8473 & 1.0078 & 1.0314 & 1.0300 & 1.0132 & 1.0507 & \textbf{0.9645}\\

\midrule

\parbox[t]{2mm}{\multirow{4}{*}{\rotatebox[origin=c]{90}{State}}}
& AETS & 0.8977 & \textbf{0.9995} & 1.0027 & 1.0033 & 1.0136 & 1.0597 & 0.9915 \\
 & AETS-BU & 1.0000 & 1.0000 & \textbf{1.0000} & 1.0000 & 1.0000 & \textbf{1.0000} & 1.0000 \\
 & DHF & 1.2194 & 1.1747 & 1.0253 & \textbf{0.9903} & \textbf{0.9775} & 1.0290 & 1.0185 \\
 & MRDLM & \textbf{0.8827} & 1.0067 & 1.0210 & 1.0239 & 1.0117 & 1.0409 & \textbf{0.9724} \\
\midrule

\parbox[t]{2mm}{\multirow{4}{*}{\rotatebox[origin=c]{90}{Store}}}
 & AETS & \textbf{0.9016} & \textbf{0.9859} & \textbf{0.9753} & \textbf{0.9855} & 0.9955 & 1.0480 & 0.9848 \\
 & AETS-BU & 1.0000 & 1.0000 & 1.0000 & 1.0000 & 1.0000 & \textbf{1.0000} & 1.0000 \\
 & DHF & 1.1930 & 1.1470 & 1.0222 & 0.9920 & \textbf{0.9775} & 1.0269 & 1.0169 \\
 & MRDLM & 0.9028 & 1.0029 & 1.0160 & 1.0199 & 1.0122 & 1.0358 & \textbf{0.9712} \\

\midrule

\parbox[t]{2mm}{\multirow{4}{*}{\rotatebox[origin=c]{90}{Category}}}
 & AETS & \textbf{0.8698} & \textbf{0.9835} & \textbf{0.9966} & 0.9977 & 1.0071 & 1.0695 & 0.9966 \\
 & AETS-BU & 1.0000 & 1.0000 & 1.0000 & 1.0000 & 1.0000 & \textbf{1.0000} & 1.0000 \\
 & DHF & 1.2516 & 1.2001 & 0.9967 & \textbf{0.9772} & \textbf{0.9652} & 1.0210 & 0.9977 \\
 & MRDLM & 0.8719 & 1.0037 & 1.0284 & 1.0361 & 1.0270 & 1.0600 & \textbf{0.9863} \\

\midrule

\parbox[t]{2mm}{\multirow{4}{*}{\rotatebox[origin=c]{90}{Department}}}
 & AETS & \textbf{0.8658} & \textbf{0.9731} & \textbf{0.9907} & 0.9992 & 1.0076 & 1.0653 & \textbf{0.9940} \\
 & AETS-BU & 1.0000 & 1.0000 & 1.0000 & 1.0000 & 1.0000 & \textbf{1.0000} & 1.0000 \\
 & DHF & 1.2404 & 1.1681 & 1.0059 & \textbf{0.9795} & \textbf{0.9726} & 1.0247 & 1.0072 \\
 & MRDLM & 0.8802 & 1.0032 & 1.0282 & 1.0378 & 1.0323 & 1.0617 & 0.9959 \\

\midrule

\parbox[t]{2mm}{\multirow{4}{*}{\rotatebox[origin=c]{90}{Sta./Cat.}}}
 & AETS & 0.9181 & 1.0010 & 1.0071 & 1.0081 & 1.0115 & 1.0591 & 1.0079 \\
 & AETS-BU & 1.0000 & \textbf{1.0000} & \textbf{1.0000} & 1.0000 & 1.0000 & \textbf{1.0000} & 1.0000 \\
 & DHF & 1.2189 & 1.1518 & 1.0285 & \textbf{0.9913} & \textbf{0.9802} & 1.0288 & 1.0106 \\
 & MRDLM & \textbf{0.9069} & 1.0070 & 1.0201 & 1.0301 & 1.0238 & 1.0484 & \textbf{0.9946} \\

\bottomrule
\end{tabular}}
\caption{M5 upper hierarchy, point forecast accuracy. Layout as per table \ref{MR-DLM-rmse}. The numbers are RMSE for the method in question, at a given layer in the hierarchy, expressed as a \% of the Bottom up aggregated base forecasts. Bold entries identify the best performing approaches.}
\label{M5-rmse-top}
\end{table}

\newpage

\begin{table}[H]
\footnotesize
\resizebox{\textwidth}{!}{%
\begin{tabular}{llrrrrrrrr}
\toprule
& & \multicolumn{7}{c}{Forecast horizon (Days)} \\
Level / Model &&              1 &              2 &              3 &              4 &              5 &              6 &              7 \\
\midrule

\parbox[t]{2mm}{\multirow{4}{*}{\rotatebox[origin=c]{90}{Sta./Dept.}}}
 & AETS & \textbf{0.9139} & \textbf{0.9998} & 1.0079 & 1.0121 & 1.0152 & 1.0575 & 1.0067 \\
 & AETS-BU & 1.0000 & 1.0000 & \textbf{1.0000} & 1.0000 & 1.0000 & \textbf{1.0000} & \textbf{1.0000} \\
 & DHF & 1.2116 & 1.1318 & 1.0340 & \textbf{0.9940} & \textbf{0.9847} & 1.0292 & 1.0142 \\
 & MRDLM & 0.9144 & 1.0067 & 1.0224 & 1.0339 & 1.0296 & 1.0511 & 1.0035 \\

\midrule

\parbox[t]{2mm}{\multirow{4}{*}{\rotatebox[origin=c]{90}{Sto./Cat.}}}
 & AETS & \textbf{0.9273} & \textbf{0.9873} & \textbf{0.9882} & 0.9973 & 1.0033 & 1.0409 & 0.9989 \\
 & AETS-BU & 1.0000 & 1.0000 & 1.0000 & 1.0000 & 1.0000 & \textbf{1.0000} & 1.0000 \\
 & DHF & 1.1885 & 1.1285 & 1.0281 & \textbf{0.9939} & \textbf{0.9821} & 1.0246 & 1.0125 \\
 & MRDLM & 0.9287 & 1.0047 & 1.0165 & 1.0255 & 1.0226 & 1.0397 & \textbf{0.9933} \\

\midrule

\parbox[t]{2mm}{\multirow{4}{*}{\rotatebox[origin=c]{90}{Sto./Dept.}}}
 & AETS & \textbf{0.9339} & \textbf{0.9924} & \textbf{0.9979} & 1.0054 & 1.0094 & 1.0405 & 1.0024 \\
 & AETS-BU & 1.0000 & 1.0000 & 1.0000 & 1.0000 & 1.0000 & \textbf{1.0000} & \textbf{1.0000} \\
 & DHF & 1.1812 & 1.1160 & 1.0351 & \textbf{0.9985} & \textbf{0.9878} & 1.0263 & 1.0177 \\
 & MRDLM & 0.9410 & 1.0080 & 1.0189 & 1.0285 & 1.0290 & 1.0431 & 1.0039 \\

\midrule

\parbox[t]{2mm}{\multirow{4}{*}{\rotatebox[origin=c]{90}{Product}}}
 & AETS & \textbf{0.9815} & 1.0176 & 1.0229 & 1.0224 & 1.0224 & 1.0205 & 1.0037 \\
 & AETS-BU & 1.0000 & 1.0000 & 1.0000 & 1.0000 & 1.0000 & 1.0000 & 1.0000 \\
 & DHF & 1.2683 & 1.1874 & 1.1330 & 1.0903 & 1.0725 & 1.1011 & 1.1383 \\
 & MRDLM & 0.9869 & \textbf{0.9916} & \textbf{0.9895} & \textbf{0.9910} & \textbf{0.9909} & \textbf{0.9970} & \textbf{0.9952} \\

\midrule

\parbox[t]{2mm}{\multirow{4}{*}{\rotatebox[origin=c]{90}{Prd./Sta.}}}
 & AETS & 0.9971 & 1.0100 & 1.0125 & 1.0118 & 1.0112 & 1.0095 & 1.0044 \\
 & AETS-BU & 1.0000 & 1.0000 & 1.0000 & 1.0000 & 1.0000 & 1.0000 & 1.0000 \\
 & DHF & 1.1591 & 1.0947 & 1.0613 & 1.0360 & 1.0258 & 1.0452 & 1.0709 \\
 & MRDLM & \textbf{0.9925} & \textbf{0.9877} & \textbf{0.9833} & \textbf{0.9831} & \textbf{0.9828} & \textbf{0.9875} & \textbf{0.9900} \\

\midrule

\parbox[t]{2mm}{\multirow{4}{*}{\rotatebox[origin=c]{90}{Prd./Sto.}}}
 & AETS & 1.0000 & 1.0000 & 1.0000 & 1.0000 & 1.0000 & 1.0000 & 1.0000 \\
 & AETS-BU & 1.0000 & 1.0000 & 1.0000 & 1.0000 & 1.0000 & 1.0000 & 1.0000 \\
 & DHF & 1.0800 & 1.0351 & 1.0134 & 0.9984 & 0.9919 & 1.0050 & 1.0226 \\
 & MRDLM & \textbf{0.9943} & \textbf{0.9835} & \textbf{0.9777} & \textbf{0.9760} & \textbf{0.9751} & \textbf{0.9800} & \textbf{0.9851} \\

\bottomrule
\end{tabular}}
\caption{M5 lower hierarchy, point forecast accuracy. Layout as per table \ref{MR-DLM-rmse}. The numbers are RMSE for the method in question, at a given layer in the hierarchy, expressed as a \% of the Bottom up aggregated base forecasts. Bold entries identify the best performing approaches.}
\label{M5-lower-rmse}
\end{table}

\newpage

\begin{table}[H]
\footnotesize
\resizebox{\textwidth}{!}{%
\begin{tabular}{llrrrrrrrr}
\toprule
& & \multicolumn{7}{c}{Forecast horizon (Days)} \\
Level / Model &&              1 &              2 &              3 &              4 &              5 &              6 &              7 \\
\midrule
\parbox[t]{2mm}{\multirow{3}{*}{\rotatebox[origin=c]{90}{Total}}}
 & AETS-BU & 1.0000 & 1.0000 & 1.0000 & 1.0000 & 1.0000 & 1.0000 & 1.0000 \\
 & MRDLM & 0.9315 & 0.8635 & \textbf{0.8369} & \textbf{0.9084} & \textbf{0.9012} & \textbf{0.8969} & 0.8838 \\
 & DHF & \textbf{0.8361} & \textbf{0.8290} & 0.8381 & 0.9104 & 0.9075 & 0.8983 & \textbf{0.8708} \\

\midrule

\parbox[t]{2mm}{\multirow{3}{*}{\rotatebox[origin=c]{90}{State}}}
 & AETS-BU & 1.0000 & 1.0000 & 1.0000 & 1.0000 & 1.0000 & 1.0000 & 1.0000 \\
 & MRDLM & 0.9596 & 0.8968 & 0.8586 & 0.8961 & 0.8958 & 0.8911 & 0.9352 \\
 & DHF & \textbf{0.8504} & \textbf{0.8368} & \textbf{0.8204} & \textbf{0.8614} & \textbf{0.8646} & \textbf{0.8541} & \textbf{0.8751} \\

\midrule

\parbox[t]{2mm}{\multirow{3}{*}{\rotatebox[origin=c]{90}{Store}}}
 & AETS-BU & 1.0000 & 1.0000 & 1.0000 & 1.0000 & 1.0000 & 1.0000 & 1.0000 \\
 & MRDLM & 0.9018 & 0.8208 & 0.8001 & 0.8242 & 0.8138 & 0.8295 & 0.8912 \\
 & DHF & \textbf{0.8164} & \textbf{0.7946} & \textbf{0.7849} & \textbf{0.8065} & \textbf{0.8044} & \textbf{0.8067} & \textbf{0.8291} \\

\midrule

\parbox[t]{2mm}{\multirow{3}{*}{\rotatebox[origin=c]{90}{Category}}}
 & AETS-BU & 1.0000 & 1.0000 & 1.0000 & 1.0000 & 1.0000 & 1.0000 & 1.0000 \\
 & MRDLM & 0.8897 & 0.8382 & \textbf{0.8219} & \textbf{0.8569} & \textbf{0.8594} & \textbf{0.8848} & 0.9090 \\
 & DHF & \textbf{0.8230} & \textbf{0.8200} & 0.8298 & 0.8679 & 0.8773 & 0.8851 & \textbf{0.8965} \\

\midrule

\parbox[t]{2mm}{\multirow{3}{*}{\rotatebox[origin=c]{90}{Department}}}
 & AETS-BU & 1.0000 & 1.0000 & 1.0000 & 1.0000 & 1.0000 & 1.0000 & 1.0000 \\
 & MRDLM & 0.9906 & 0.8751 & 0.8468 & 0.9019 & 0.9375 & 0.9489 & 0.9959 \\
 & DHF & \textbf{0.8225} & \textbf{0.8151} & \textbf{0.8178} & \textbf{0.8693} & \textbf{0.8841} & \textbf{0.8605} & \textbf{0.8852} \\

\midrule

\parbox[t]{2mm}{\multirow{3}{*}{\rotatebox[origin=c]{90}{Sta./Cat.}}}
 & AETS-BU & 1.0000 & 1.0000 & 1.0000 & 1.0000 & 1.0000 & 1.0000 & 1.0000 \\
 & MRDLM & 0.9464 & 0.8918 & 0.8454 & 0.8702 & 0.8716 & 0.9425 & 0.9622 \\
 & DHF & \textbf{0.8262} & \textbf{0.8262} & \textbf{0.8162} & \textbf{0.8450} & \textbf{0.8467} & \textbf{0.8733} & \textbf{0.8880} \\

\bottomrule
\end{tabular}}
\caption{M5 upper hierarchy, probablistic forecast accuracy. Layout as per table \ref{MR-DLM-rmse}. The numbers are scores for the method in question, at a given layer in the hierarchy, expressed as a \% of the Bottom up aggregated base forecasts. Bold entries identify the best performing approaches.}
\label{M5-score-top}
\end{table}

\newpage
\begin{table}[H]
\footnotesize
\resizebox{\textwidth}{!}{%
\begin{tabular}{llrrrrrrrr}
\toprule
& & \multicolumn{7}{c}{Forecast horizon (Days)} \\
Level / Model &&              1 &              2 &              3 &              4 &              5 &              6 &              7 \\
\midrule

\parbox[t]{2mm}{\multirow{3}{*}{\rotatebox[origin=c]{90}{Sta./Dept.}}}
 & AETS-BU & 1.0000 & 1.0000 & 1.0000 & 1.0000 & 1.0000 & 1.0000 & 1.0000 \\
 & MRDLM & 0.9525 & 0.8732 & 0.8398 & 0.8775 & 0.8826 & 0.9283 & 0.9621 \\
 & DHF & \textbf{0.8176} & \textbf{0.8095} & \textbf{0.8052} & \textbf{0.8429} & \textbf{0.8430} & \textbf{0.8481} & \textbf{0.8697} \\

\midrule

\parbox[t]{2mm}{\multirow{3}{*}{\rotatebox[origin=c]{90}{Sto./Cat.}}}
 & AETS-BU & 1.0000 & 1.0000 & 1.0000 & 1.0000 & 1.0000 & 1.0000 & 1.0000 \\
 & MRDLM & 0.8946 & 0.8322 & 0.8063 & 0.8236 & 0.8143 & 0.8637 & 0.9031 \\
 & DHF & \textbf{0.8110} & \textbf{0.7968} & \textbf{0.7888} & \textbf{0.8060} & \textbf{0.8029} & \textbf{0.8223} & \textbf{0.8399} \\

\midrule

\parbox[t]{2mm}{\multirow{3}{*}{\rotatebox[origin=c]{90}{Sto./Dept.}}}
 & AETS-BU & 1.0000 & 1.0000 & 1.0000 & 1.0000 & 1.0000 & 1.0000 & 1.0000 \\
 & MRDLM & 0.9209 & 0.8396 & 0.8126 & 0.8411 & 0.8248 & 0.8640 & 0.9142 \\
 & DHF & \textbf{0.8108} & \textbf{0.7945} & \textbf{0.7909} & \textbf{0.8155} & \textbf{0.8082} & \textbf{0.8188} & \textbf{0.8380} \\

\midrule

\parbox[t]{2mm}{\multirow{3}{*}{\rotatebox[origin=c]{90}{Product}}}
 & AETS-BU & 1.0000 & 1.0000 & 1.0000 & 1.0000 & 1.0000 & 1.0000 & 1.0000 \\
 & MRDLM & 1.3189 & 1.0815 & 1.0328 & 1.0266 & 0.9924 & 1.0264 & 1.0387 \\
 & DHF & \textbf{0.8824} & \textbf{0.8140} & \textbf{0.8016} & \textbf{0.7971} & \textbf{0.7752} & \textbf{0.8014} & \textbf{0.7938} \\

\midrule

\parbox[t]{2mm}{\multirow{3}{*}{\rotatebox[origin=c]{90}{Prd./Sta.}}}
 & AETS-BU & 1.0000 & 1.0000 & 1.0000 & 1.0000 & 1.0000 & 1.0000 & 1.0000 \\
 & MRDLM & 1.1579 & 1.0227 & 1.0057 & 1.0068 & 1.0063 & 1.0147 & 1.0211 \\
 & DHF & \textbf{0.8534} & \textbf{0.8072} & \textbf{0.7837} & \textbf{0.7932} & \textbf{0.7991} & \textbf{0.7927} & \textbf{0.7888} \\

\midrule

\parbox[t]{2mm}{\multirow{3}{*}{\rotatebox[origin=c]{90}{Prd./Sto.}}}
 & AETS-BU & 1.0000 & 1.0000 & 1.0000 & 1.0000 & 1.0000 & 1.0000 & 1.0000 \\
 & MRDLM & 1.0275 & 1.0156 & 1.0001 & 0.9945 & 0.9978 & 0.9969 & 1.0039 \\
 & DHF & \textbf{0.7898} & \textbf{0.8022} & \textbf{0.7902} & \textbf{0.7883} & \textbf{0.7914} & \textbf{0.7869} & \textbf{0.7816} \\

\bottomrule
\end{tabular}}
\caption{M5 lower hierarchy, probablistic forecast accuracy. Layout as per table \ref{MR-DLM-rmse}. The numbers are scores for the method in question, at a given layer in the hierarchy, expressed as a \% of the Bottom up aggregated base forecasts. Bold entries identify the best performing approaches.}
\label{M5-lower-score}
\end{table}

\newpage

\section{Appendix E - SVD Forward Filtering for the DLM}

The following is based on \cite{Petris2009-eq} and \cite{Willard2020-no}. For the DLM:

\begin{align*}
    y_t &= F'_t \theta_t + \epsilon_t & \epsilon &\sim N(0,V_t)\\
    \theta_t &= G_t \theta_{t-1} + \upsilon_t & \upsilon_t &\sim N(0,W_t)
\end{align*}

With priors at time t:

\begin{align*}
    \theta_t &\sim N(a_t,R_t)\\
    y_t &\sim N(f_t,Q_t)
\end{align*}

The 1 step ahead forecast at time t is:

\begin{align*}
    a_t &= G_t m_{t-1} & R_t &= G_tC_{t-1}G'_t + W_t\\
    f_t &= F'_ta_t & Q_t &= F'_tR_tF_t + V_t
\end{align*}

And the posterior on observing $y_t$ is:

\begin{align*}
    \theta_t &\sim N(m_t,C_t)\\
    m_t &= a_t + R_tF_tQ^{-1}_t (f_t-y_t)\\
    C_t &= R_t-R_tF_tQ^{-1}_tF'_tR_t
\end{align*}

Recall that the SVD of a matrix $M$ is given by $M=U_MD_MV_M'$.  If $M$ is symetric then $M = U_MD_MU'_M$ and its matrix square root is $N_M = S_MU'_M$ where $S_M=D^{1/2}_M$ so that $N'_MN_M=M$. Write:
\begin{align*}
    N_{R_{t}}'N_{R_{t}} &= R_t = G_tC_{t-1}G'_t+W_t\\ 
    \text{where }
    N_{R_{t}} & = \left(\begin{matrix} S_{C_{t-1}}U'_{C_{t-1}}G'_t \\ N_{{W_t}} \end{matrix}\right) 
\end{align*}

So that the matrix square root of $R_t$ is expressed in terms of the SVD of $C_{t-1}$ and the matrix square root of $W_t$. Using the Woodbury identity we can write $C^{-1}_t = F_tV^{-1}_tF'_t + R^{-1}_t$. Introducing an additional term in $U_{R_{t}}$, we have $C^{-1}_t = U'_{R_{t}}F_tV^{-1}_tF'_tU_{R_{t}} + U'_{R_{t}}R^{-1}_tU_{R_{t}}$, the matrix square root of which can be written as:

\begin{align*}
    N_{C^{-1}_{t}} & = \left(\begin{matrix} N_{V^{-1}_{t}}F'_tU_{R_{t}} \\
    S^{-1}_{{R_t}} \end{matrix}\right) 
\end{align*}

Taking the SVD of $N_{C^{-1}_{t}}$, the key components of the SVD of $C_t$ are then given by $U_{C_{t}} = U_{R_{t}}V_{N_{C^{-1}_{t}}}$ and $S_{C_{t}} = D^{-1}_{N_{C^{-1}_{t-1}}}$.

\newpage
\section{Appendix F - Dynamic combination regressions}

From the main text above, the combination regression, assuming the regressors in $\mathbf{F}_t$ are known, written using only the base series is:

\begin{align*}
\mathbf{b}_{t}&=\mathbf{F}'_{t} \boldsymbol{\theta}_{t}+\boldsymbol{\varepsilon}_{t} &
\quad \boldsymbol{\varepsilon}_{t} &\sim N\left(0, \mathbf{V}_{t}\right) \\ \boldsymbol{\theta}_{i,t}&=\boldsymbol{\theta}_{t-1}+\boldsymbol{\omega_{t}} &\quad \boldsymbol{\omega}_{t} &\sim N\left(0, \mathbf{W}_{t})\right.
\end{align*}

Where $\mathbf{F}'_t$ is an $n_b \times (n_b \times k)$ block diagonal matrix with the regressors for each base series as diagonal blocks, and $\boldsymbol{\theta}_t$ is an $(n_b \times k)$ vector comprising the regression weights.

In order to estimate $\boldsymbol{\theta}_t$ accounting for the aggregates we pre-multiply by $\mathbf{S}$ and write:

\begin{align}
\mathbf{S}\mathbf{b}_{t}&=\mathbf{S}\mathbf{F}'_{t} \boldsymbol{\theta}_{t}+\mathbf{S}\boldsymbol{\varepsilon}_{t}  & \quad  \mathbf{S} \boldsymbol{\varepsilon_{t}} &\sim N\left(0, \mathbf{S}\mathbf{V}_{t}\mathbf{S}'\right) \\  \boldsymbol{\theta}_{t}&=\boldsymbol{\theta}_{t-1}+\boldsymbol{\omega_{t}} &\quad \boldsymbol{\omega}_{t} &\sim N\left(0, \mathbf{W}_{t})\right.
\end{align}

Estimation of these regressions is greatly simplified by assuming the variance $\mathbf{V}_t$ is known and equal to the variance / covariance matrix estimated via the MRDLM model, $\hat{\mathbf{Q}}_t$. Such assumptions, using pre-estimated covariance matrices has a long history in Bayesian time series analysis, and is used for example in Vector Auto Regressions utilising the Minnesota prior (\cite{Doan1984-ju}, \cite{Litterman1986-lq}). Updating and forecasting in this model follow standard DLM equations assuming known variance and deterministic regressors, with $\mathbf{S}\mathbf{b}_{t}$ replacing the vector of observations, and $\mathbf{S}\mathbf{F}'_{t}$ as the matrix $\mathbf{F}'_t$ in usual DLM notation. The model is:

\begin{align*}
    \mathbf{Sb}_t &= \mathbf{SF}'_t \boldsymbol{\theta}_{t} + \mathbf{S} \boldsymbol{\epsilon}_t & \epsilon_t &\sim N(0,\mathbf{V}_t)\\
    \boldsymbol{\theta}_{t} &= \boldsymbol{\theta}_{t-1} +\boldsymbol{\omega_{t}} & \boldsymbol{\omega}_{t} &\sim N\left(0, \mathbf{W}_{t})\right.
\end{align*}

With $F_t$ representing the dis-aggregated base forecasts for the set of base series and priors on the state vector $\boldsymbol{\theta}_t \sim [m_{t-1},C_{t-1}]$ at time t, the moments of the state vector and the 1 step ahead forecast at time t is:

\begin{align*}
    \boldsymbol{\theta}_t &\sim  \left[\mathbf{a}_t,\mathbf{R}_t\right]\\
    \mathbf{y}_t &\sim \left[\mathbf{f}_t,\mathbf{Q}_t\right]
\end{align*}

With:

\begin{align*}
    \mathbf{a}_t &= \mathbf{m}_{t-1} & \mathbf{R}_t &= \mathbf{C}_{t-1} + \mathbf{W}_t\\
    \mathbf{f}_t &= \mathbf{SF}'_t \mathbf{a}_t & \mathbf{Q}_t &= \mathbf{SF}'_t\mathbf{R}_t\mathbf{F}_t\mathbf{S}' + \mathbf{SV}_t\mathbf{S}'
\end{align*}

$\mathbf{V}_t$ assumed known. The posterior state vector after observing $y_t$ is:

\begin{align*}
    \boldsymbol{\theta}_t &\sim N\mathbf{(m}_t,\mathbf{C}_t)\\
    \mathbf{m}_t &= \mathbf{a}_t + \mathbf{R}_t\mathbf{F}_t\mathbf{S'Q}^{-1}_t (\mathbf{f}_t-\mathbf{y}_t)\\
    \mathbf{C}_t &= \mathbf{R}_t-\mathbf{R}_t\mathbf{F}_t\mathbf{S'Q}^{-1}_t\mathbf{F}'_t\mathbf{SR}'_t
\end{align*}

Where the dis-aggregated forecasts are treated as random quantities, we modify the preceding analysis as follows.
For a single series $i$, the DLM might be written as $b_{it} = \mathbf{F}_{it}\boldsymbol{\theta}_{it} + e_{it}$, $\boldsymbol{\theta}_{it} = \boldsymbol{\theta}_{i,t-1} + \boldsymbol{\omega}_{it}$, $e_{it} \sim [0,s_{it}]$ and $\mathbf{w}_{it} \sim [0,\mathbf{W}_{it}]$, where $\boldsymbol{F}_{it} \sim [\mathbf{h}_{it},\mathbf{H}_{it}]$. The DLM updating equations would then be modified so that the one period ahead forecast of $b_{it}$ is $f_{it} = \mathbf{h}'_{it} \mathbf{a}_{it}$ with variance $q_{it} = \mathbf{h}_{it}' \mathbf{R}_{it} \mathbf{h}_{it} + \mathbf{a}'_{it} \mathbf{H}_{it} \mathbf{a}_{it} + tr(\mathbf{R}_{it}\mathbf{H}_{it}) + s_{it}$ with $tr$ denoting the trace operator. For the univariate case, because we treat the base forecasts as a collection of univariate series, we do not have available the full covariance matrix of the propagated forecasts $\mathbf{H}_{it}$, so we replace this by a diagonal matrix of the estimated forecast variances from (\ref{diaggregation}).

Moving to the multivariate case, we start with each $\boldsymbol{F}_{it} \sim [\mathbf{h}'_{it},\mathbf{H}_{it}]$ where each $\mathbf{H}_{it}$ is a $k \times k$ diagonal matrix of the variances of the dis-aggregated forecasts applicable to base series $i$. Again, because we treat the base forecasts as univariate, we face a similar problem to that described above, with the additional complication of needing to account for the covariance between regressors of the various base series (which are in theory available from our analysis above). We again proceed by approximating $\mathbf{H}_t$ with a block diagonal matrix of the variances of the individual dis-aggregated forecasts. We then construct $\mathbf{F}_t$ as the $(n_b \times k) \times n_b$ block diagonal matrix:

\begin{align}
    \mathbf{F}_t = \left[
    \begin{matrix}    
        \mathbf{h}'_{1t} &  &  &  &  \\
        & ... & & & \\
        & & \mathbf{h}'_{it}  &  & \\
        & & & ... & \\
        & & & & \mathbf{h}'_{n_{b}t}  & \\
    \end{matrix}
    \right]
\end{align}

Where the $(n_b \times k)$ terms have diagonal covariance matrix:

\begin{align}
    \mathbf{H}_t = \left[
    \begin{matrix}    
        \mathbf{H}'_{1t} &  &  &  &  \\
        & ... & & & \\
        & & \mathbf{H}'_{it}  &  & \\
        & & & ... & \\
        & & & & \mathbf{H}'_{n_{b}t}  & \\
    \end{matrix}
    \right]
\end{align}

We write the DLM as $\mathbf{Sb}_{t} = \mathbf{SF}_{t}'\boldsymbol{\theta}_{t} + \mathbf{Se}_{t}$, $\boldsymbol{\theta}_{t} = \boldsymbol{\theta}_{t-1} + \boldsymbol{\omega}_{t}$, $\mathbf{e}_{t} \sim [0,\mathbf{V}_{t}]$ and $\boldsymbol{\omega}_{t} \sim [0,\mathbf{W}_{t}]$.

The DLM updating equations are again modified so that the one period ahead forecast of $\mathbf{y}_{t}$ is $\mathbf{Sf}_{t} = \mathbf{SF}'_t \mathbf{m}_{t}$ with variance $\mathbf{SQS}'_t = \mathbf{S} \left[ \mathbf{F}'_t \mathbf{R}_{t} \mathbf{F}_t + \mathbf{m}'_{t} \mathbf{H}_{t} \mathbf{m}_{t} + tr(\mathbf{R}_{t} \mathbf{H}_t) + \mathbf{V}_t \right] \mathbf{S}'$. 

When constructing multi step ahead forecasts we estimate the revised forecast error variance / covariance matrix given the exogenous forecasts using a linear bayes estimator $\bar{\mathbf{Q}}_t - \mathbf{m}'_t\mathbf{H}_t\mathbf{m}_t$ where $\mathbf{H}_t$ is a block diagonal matrix of the estimated variances of the dis-aggregated base forecasts.

The reconciled forecast distributions of the base series are then:

\begin{align}
    \tilde{\mathbf{b}}_t \sim \left[\tilde{\mathbf{f}}_t,\tilde{\mathbf{Q}}_t\right]
\end{align}
\begin{align}
    \tilde{\mathbf{f}}_t & = \mathbf{F}_t \boldsymbol{\theta}_t & 
    \tilde{\mathbf{Q}}_t=&\frac{\nu_t}{\nu_t-2} \left[ \bar{\mathbf{Q}}_t - \mathbf{m}_t \mathbf{H}_{t} \mathbf{m}'_{t} + \tilde{\mathbf{F}}_{t}' \mathbf{R}_{t}' \mathbf{F}_{t} + tr(\mathbf{R}_{t}\mathbf{H}_{t}) \right] + \mathbf{m}_{t}' \mathbf{H}_{t} \mathbf{m}_{t}
\end{align}

\newpage
\section{Appendix G - Hierarchical Prior Estimation}

From the main text, the observation equation is:

\begin{align}
    \mathbf{y}_t = \mathbf{Sb}_{t}&=\mathbf{SF}_{bt} \boldsymbol{\theta}_{bt}+\mathbf{Sv}_{bt} &
    \quad \mathbf{v}_{bt} &\sim N\left(0, \mathbf{V}_{bt}\right)
\end{align}

The vector of regression parameters for each base series $\boldsymbol{\theta}_{bt}$ is related to a shared, hierarchy wide set of parameters $\boldsymbol{\theta}_{ht}$ via a structural equation:

\begin{align}
    \label{hier-comb-regression-struct-eqn}
    \boldsymbol{\theta}_{bt}&=\mathbf{F}_{ht} \boldsymbol{\theta}_{ht}+v_{ht} &\quad v_{t} &\sim N\left(0, \mathbf{V}_{ht})\right.
\end{align}

The hyper parameter $\mathbf{V}_{ht}$ controls the degree to which series level combination weights differ from their hierarchy wide estimates. These shared parameters then evolve over time as:

\begin{align}
\boldsymbol{\theta}_{ht}&=\boldsymbol{\theta}_{h,t-1}+\boldsymbol{\omega_{t}} &\quad \boldsymbol{\omega}_{t} &\sim N\left(0, \mathbf{W}_{t})\right.
\end{align}.

Write:

\begin{align*}
    V_{ht} &= I_{n_{b}} \oplus V & \beta_{ti} & \sim N(\mu_{it},V)
\end{align*}

Where $\mathbf{F}_{1t} = diag(\mathbf{h}'_{t1}...\mathbf{h}'_{tk})$ are the regressors (dis-aggregated forecasts from different levels as above), $\boldsymbol{\theta_{bt}}$ the bottom level series regression coefficients, $\mathbf{V}_{bt}$ is the observation variance / covariance matrix (which we again approximate with $\bar{mathbf{Q}}_t$) and $F_{ht} = 1_n \oplus I_{n_{b}}$. Inference proceeds broadly as set out in \cite{Gamerman1993-xw}, but with the additional terms in the expression for $Q_t$ due to the uncertainty in the regressors as set out above.

\begin{enumerate}
    \item Prior distributions at time t:
    $\boldsymbol\theta_{ht} \sim \left[\mathbf{a}_{ht},\mathbf{R}_{ht}\right]$, with $\mathbf{a}_{ht} = \mathbf{m}_{ht}$ and $\mathbf{R}_{ht} = \mathbf{C}_{h,t-1} +  W_t$, and $\boldsymbol{\theta}_{bt} \sim \left[\mathbf{a}_{bt},\mathbf{R}_{bt}\right]$, with $\mathbf{a}_{bt} = \mathbf{F}_{bt}\mathbf{a}_{bt}$ and $\mathbf{R}_{bt} = \mathbf{F}_{ht}\mathbf{C}_{h,t-1}\mathbf{F}'_{ht} +  \mathbf{V}_{ht}$. 
    \item Predictive distributions (one step ahead):
    $\mathbf{b}_t|D_{t-1} \sim N(\mathbf{f}_t,\mathbf{Q}_t)$ with $\mathbf{f}_t = \mathbf{F}_{bt}\mathbf{a}_{bt}$ and $\mathbf{Q}_t = ( \mathbf{F}'_{bt} \mathbf{R}_{bt} \mathbf{F}_{bt} + \mathbf{m}'_{bt} \mathbf{H}_{t} \mathbf{m}_{bt} + tr(\mathbf{R}_{bt} \mathbf{H}_{t}) + \mathbf{V}_{bt} )$
    \item Posterior Distributions at time t: $\boldsymbol{\theta}_{bt} \sim N(\mathbf{m}_{bt},\mathbf{C}_{bt})$, $\boldsymbol{\theta}_{ht} \sim \left[ \mathbf{m}_{ht},\mathbf{C}_{ht} \right]$.
    Write $\mathbf{E}_{0,b,t} = \mathbf{F}_{bt}$, $\mathbf{E}_{0,h,t} = \mathbf{F}_{bt}\mathbf{F}_{ht}$. Then $\mathbf{m}_{it} = \mathbf{a}_{it} + \mathbf{X}_{it}(\mathbf{SQS}')^{-1}(\mathbf{y}_t-\mathbf{f}_t)$, $\mathbf{C}_{it} = \mathbf{R}_{it} - \mathbf{X}_{it}(\mathbf{SQS}')^{-1}_t\mathbf{X}'_{it}$ with $\mathbf{X}_{it}=\mathbf{R}_{it}\mathbf{E}'_{0it}$ for hierarchical levels $i = \{h,b\}$. Note that we use an $\mathbf{X}$ instead of the $\mathbf{S}$ used in \cite{Gamerman1993-xw}.
\end{enumerate}    

We produce multi step ahead forecasts as set out above in Appendix F.

\clearpage
\bibliographystyle{elsarticle-harv}  
\bibliography{references}

\end{document}